\documentclass[%
reprint,
aps,
]{revtex4-2}

\usepackage[utf8x]{inputenc}
\usepackage{graphicx}

\usepackage{amsmath}
\usepackage{amssymb}
\usepackage{siunitx}
\usepackage{mathtools}

\usepackage{hyperref}
\usepackage[pdftex,dvipsnames]{xcolor}
\hypersetup{frenchlinks = true,
	colorlinks,
	linkcolor={red!50!black},
	citecolor={blue!50!black},
	urlcolor={blue}
}

\usepackage{lipsum}
\usepackage{braket}
\usepackage{enumitem}

\usepackage{lmodern}
\usepackage{slantsc}

\usepackage{xargs}                      

\paperwidth=\dimexpr \paperwidth + 6cm\relax 
\oddsidemargin=\dimexpr\oddsidemargin + 3cm\relax
\evensidemargin=\dimexpr\evensidemargin + 3cm\relax
\marginparwidth=\dimexpr \marginparwidth + 3cm\relax

\def\mat#1{\mathbf{#1}}
\def\op#1{\hat{#1}}
\def\st#1{_{\mathrm{#1}}}
\def\ut#1{^{\mathrm{#1}}}

\usepackage{bm}
\renewcommand\vec{\bm}

\renewcommand{\d}[0]{\mathrm{d}}

\newcommand{\crab}{\textsc{crab}}
\newcommand{\bfgs}{\textsc{bfgs}}
\newcommand{\dcrab}{d\textsc{crab}}

\newcommand{\U}[0]{\op{\mathcal{U}}}
\renewcommand{\H}[0]{\op{H}}
\newcommand{\h}[0]{\op{h}}
\newcommand{\Hd}[0]{\H^{d}}
\newcommand{\Hc}[0]{\H^{c}}
\newcommand{\psio}[0]{\psi\st{ini}}
\newcommand{\dt}[0]{\delta t}
\newcommand{\F}[0]{F}
\newcommand{\overlap}[0]{o}
\newcommand{\nt}[0]{N_{t}}
\newcommand{\nsites}[0]{N_{s}}

\renewcommand{\Re}{\mathrm{Re}}

\newcommand{\SF}{\ket{\mathrm{SF}}}
\newcommand{\Mott}{\ket{\mathrm{Mott}}}

\newcommand{\mps}{\textsc{mps}}

\usepackage{xspace}

\newcommand{\dmrg}{\textsc{dmrg}\xspace}

\newcommand{\g}{g\st{3D}}

\newcommand{\Hdbar}{\hat{\oline{H}}_n~\!\!\!\!\!^d}

\newcommand{\psitgt}[1][]{%
	\ifthenelse{\equal{#1}{}}{\psi_{\mathrm{tgt}}}{\psi_{\mathrm{tgt,#1}}}%
}

\newcommand{\tqsl}[1][]{%
	\ifthenelse{\equal{#1}{}}{T\st{min}^F}{T\st{min}^{F=#1}}%
}

\newcommand*\oline[1]{\overline{#1}}

\usepackage{pgfplots}

\begin{document}
\title{Achieving fast high-fidelity optimal control of many-body quantum dynamics}
	\author{Jesper Hasseriis Mohr Jensen$^1$} \email{jhasseriis@phys.au.dk}
	\author{Frederik Skovbo Møller$^{1,2}$}
	\author{Jens Jakob Sørensen$^1$}
	\author{Jacob Friis Sherson$^1$}
	\email{sherson@mgmt.au.dk}
		\affiliation{%
		$^1$ Department of Physics and Astronomy, Aarhus University, Ny Munkegade 120, 8000 Aarhus C, Denmark
	}
	\affiliation{%
		$^2$ Vienna Center for Quantum Science and Technology, Atominstitut, TU Wien, Stadionallee 2, 1020 Vienna, Austria
	}

\begin{abstract}
	We demonstrate the efficiency of a recent exact-gradient optimal control methodology by applying it to a challenging many-body problem, crossing the superfluid to Mott-insulator phase transition in the Bose-Hubbard model. The system size necessitates a matrix product state representation and this seamlessly integrates with the requirements of the algorithm. We observe fidelities in the range 0.99-0.9999 with associated minimal process duration estimates displaying an exponential fidelity-duration trade-off across several orders of magnitude. The corresponding optimal solutions are characterized in terms of a predominantly linear sweep across the critical point followed by bang-bang-like structure. This is quite different from the smooth and monotonic solutions identified by earlier gradient-free optimizations which are hampered in locating the higher complexity protocols in the regime of high-fidelities at low process durations. Overall, the comparison suggests significant methodological improvements also for many-body systems in the ideal open-loop setting. Acknowledging that idealized open-loop control may deteriorate in actual experiments, we discuss the merits of using such an approach in combination with closed-loop control --- in particular, high-fidelity physical insights extracted with the former can be used to formulate practical, low-dimensional search spaces for the latter. 
\end{abstract}

\maketitle

\section{Introduction}
With experimental and theoretical advances in the preparation and engineering of quantum mechanical systems, precise manipulation of fragile quantum systems has become increasingly important \cite{acin2018quantum}. 
To this end, \textit{quantum optimal control} \cite{glaser2015training} is a particularly successful tool for designing controls that implement desired physical transformations with wide applications in numerous research areas, such as
superconducting qubits \cite{motzoi2009simple,egger2013optimized,goerz2017charting,montangero2018introduction,mogens2020global},
nuclear magnetic resonance systems \cite{kehlet2004improving,khaneja2005optimal,nielsen2007optimal,kallies2018cooperative,sorensen2020optimization}, 
nitrogen vacancy centers \cite{scheuer2014precise,dolde2014high,waldherr2014quantum,chou2015optimal}, 
cold molecules \cite{koch2004stabilization,koch2006making,de2011optimal,tibbetts2013optimal}, and
cold atoms \cite{doria2011optimal,van2016optimal,mundt2009optimalcontrol, jager2013optimal,cui2017optimal,patsch2018fast,omran2019generation,larrouy2020fast},
to name a few.
On the theoretical side, open-loop optimal control design can be considered the union of numerical simulation and optimization methodologies.
Therefore, with growing Hilbert space sizes such as in many-body contexts \cite{georgescu2014quantum,blatt2012quantum, sachdev2011quantumphasetransitions,amico2008entanglement, kollath2007quench,eisert2015quantum,frerot2018quantum,garbe2020critical,osterloh2002scaling,de2018genuine,valdez2017quantifying} the performance capacity of both these components must be streamlined to succeed in finite time within this paradigm. 

Simulating very high-dimensional many-body systems exactly requires an exponential amount of memory and computation time \cite{poulin2011quantum}. 
Current techniques for breaking this \textit{curse of dimensionality} finds recourse in tensor networks ansätze and, in the case of 1D systems, the appropriate structures are \textit{matrix product states} (\mps). The entanglement entropy for these systems exhibit a constant area scaling law with the number of particles and can therefore be effectively simulated classically with subexponential resources \cite{poulin2011quantum,vidal2004efficient,vidal2003efficient,lloyd2014information,montangero2018introduction}.  

Among the significant number of nonlinear optimization methodologies 
\cite{khaneja2005optimal,de2011second,machnes2011comparing,floether2012robust,goodwin2015auxiliary,goodwin2016modified,machnes2018tunable,sorensen2018quantum,sorensen2020optimization,tannor1992control,palao2002quantum,schirmer2011efficient,caneva2011chopped,doria2011optimal,van2016optimal,li2018global},
it is conventional understanding \cite{nocedal2006numerical,sorensen2018quantum, machnes2018tunable} that derivative-based local search techniques is the best way to identify local optima for any given optimization objective $J$. 
Coupled with an appropriate multistarting scheme \cite{ugray2007multistart}, global optima may be uncovered. 
In addition to the local gradient $\nabla J$ it is well-known that also incorporating curvature information through the local Hessian $\nabla^2 J$ is essential for the convergence rates of local optimization. 
The theoretical ideal is to include both exactly, but in practice the full Hessian is often prohibitively expensive to calculate.
It is instead standard practice to apply e.g. \textsc{bfgs} schemes to progressively construct a Hessian approximation from successive gradients \cite{nocedal2006numerical}. If the individual gradients are inexact these errors will accumulate in each iteration and make the Hessian approximation unreliable as the optimization progresses. This will in turn manifest as increasingly poor update directions that ultimately leads the 
optimization irreparably astray and significantly hamper the discovery of local optima as discussed in e.g. Refs.~\cite{de2011second,jensen2021approximate}.
Hence exact gradients are central to local search paradigms.
Applying these in very high-dimensional Hilbert spaces, however, has been a major obstacle because the currently known exact gradient element calculations become ``extremely resource consuming, if not impossible'' \cite{doria2011optimal} due to their unfavorable scaling with the Hilbert space dimensionality $D_\mathcal{H}$.
Specifically, their evaluation requires either full diagonalization \cite{dalgaard2020hessian} or exponentiation \cite{goodwin2016modified} of matrices that are atleast of size $D_\mathcal{H}\times D_\mathcal{H}$ potentially followed by a recursive commutator series summation \cite{de2011second,jensen2021approximate}. 
Therefore, control problems in e.g. the many-body limit have either been approached using derivative-free methods \cite{doria2011optimal,van2016optimal} or, contemporarily to this work, inexact first-order gradients \cite{quinones2019tensor} which are prone to the reduced overall optimization capabilities described above.

In this paper we apply a new exact derivative methodology, introduced in our recent parallel work \cite{jensen2021approximate} and reviewed in Sec.~\ref{sec:statetransfer}, to a paradigmatic problem in the complex many-body regime. 
Briefly, the methodology circumvents the computational bottlenecks mentioned above. 
This is achieved by 1) applying Trotterization schemes, and 2) representing the problem in a diagonal basis for the control Hamiltonian. 
Explicitly including these effects in the analytical derivations leads to very simple forms for the gradient depending on the specific Trotter scheme. 
The expressions are exact with only a first-order term as all higher-order correction terms vanish and the only computational effort lies in time evolving an auxiliary state in addition to the usual quantum state. 
That is, these exact gradient calculations scale only with the time it takes to solve the quantum dynamics which is the fundamental operation of any numerical optimal control algorithm. 
It is of particular convenience that both 1) and 2) are by themselves common and independently implemented in the general context of numerical simulations. 
For example, a large family of algorithms for performing time evolution of the matrix product states considered here rely on Trotterization \cite{paeckel2019time}. 
Our exact derivative methodology thus meshes quite naturally with established standard numerical techniques. 

The central ideas and derivations in Ref.~\cite{jensen2021approximate} are valid for all unitary control tasks. 
For concreteness, the special case of maximizing state transfer fidelity in Hilbert spaces of dimension $D_\mathcal{H}=2-100$ was considered there. 
For fixed $D_\mathcal{H}$ we demonstrated orders of magnitude improvement in terms of computational speed-up relative to similar exact gradients and that the gained relative speed-up is exponential as a function of $D_\mathcal{H}$. 
It was also demonstrated that inexact gradients lead to the expected poor performance in terms of achievable fidelity. 

Here we demonstrate that the new techniques remain viable for fidelity requirements above 0.99 for many-body systems with $D_\mathcal{H}\sim10^{11}$ where other currently known exact gradient approaches are computationally prohibited. 
This evidences the possibility for significantly enhanced optimal open-loop control  also over very high-dimensional Hilbert spaces operating in the fast high-fidelity regime.

Having showcased an advance in open-loop control capabilities, we finally turn to the broader context of quantum optimal control. 
Outlining first the benefits of open- and closed-loop methodologies separately, we give our perspective on how the role played by the open-loop component in a ``unified-loop'' may remain useful in the future.

\section{Many-Body State Transfer}
\label{sec:statetransfer}

\subsection{Bose-Hubbard model}
\label{sec:bhmps}
As a challenging representative example from the class of many-body problems, we examine the superfluid-Mott insulator phase transition in the one-dimensional Bose-Hubbard model. 
This model describes the physics of $N_p$ interacting spinless bosons in a lattice with $N_s$ sites by the Hamiltonian  \cite{greiner2002quantum,lewenstein2007ultracold,georgescu2014quantum} 
\begin{align}
&\op H\st{SI} 
= \H^{J_x} + \H^U = J_x \sum_{i=1}^{N_s-1} \h^{J_x}_{[i,i+1]} + \frac{U}{2}\sum_{i=1}^{N_s} \h^U_{[i]}, \label{eq:H} \\
&\h^{J_x}_{[i,i+1]} =  -(\op a_{i+1}^\dagger \op  a_{i} + \mathrm{h.c.}), \hspace{1cm} \h^U_{[i]} = \op n_i (\op n_i -1).
\end{align}
The operators $\op a_i^\dagger$ and $\op a_i$ are the bosonic creation and annihilation operators for site $i$, respectively, while $\op n_i = \op a_i^\dagger \op a_i$ counts the number of particles occupying the site, $\op n_i \ket{n_1, \dots, n_i, \dots , n_{N_s}} = n_i \ket{n_1, \dots, n_i, \dots , n_{N_s}}$. We assume a fixed number of particles $N_p = \sum_{i=1}^{N_s} n_i$ and unit filling $N_s/N_p = 1$. 

The energy $J_x$ is associated with the hopping/tunneling operator $ \h^{J_x}_{[i,i+1]}$ along the $x$ direction, and $U$ is the energy associated with the on-site interaction operator $\h^U_{[i]}$. 
The ratio $U/J_x$ characterizes the quantum phase of the system. 
We associate with $J_x \gg U$ the superfluid phase in which the ground state $\SF$ is a delocalized particle distribution across the lattice with sizeable site occupation variance.
The Mott insulator phase is conversely associated with $U \gg J_x$ and the ground state $\Mott = \ket{1,1,\dots,1}$ is a single Fock state component with unit occupancy on each site in the thermodynamic limit. 

We seek to dynamically connect the ground states, $\SF\rightarrow\Mott$, on opposing sides of the critical point by controlling the time-dependent ratio of on-site interaction- and tunneling energies $u(t) = U(t)/J_x(t)$. 
In particular, we are interested in estimating the lowest possible transfer duration $T$ consistent with high-fidelity requirements, e.g. $F(T)\geq 0.99$, yielding empirical estimates for the minimal process duration $T\st{min}^F$.
To this end we turn to numerical quantum optimal control techniques as described in Sec.~\ref{sec:exactgradient}. 
A particular difficulty in driving the transition is that the energy excitation spectrum becomes approximately gapless 
(exactly gapless in the thermodynamic limit) in the superfluid limit.
This implies very long (diverging) adiabatic time scales 
for crossing the critical point of the phase transition. 
Thus, residual population/defects may become pinned which can be estimated either by Kibble-Zurek theory or as a cascade of independent LZ avoided crossing transitions \cite{damski2005simplest,santoro2002theory,caneva2011speeding, caneva2009optimal}. 
\\

One possible physical context for desiring such a fast and precise transfer lies in neutral atoms trapped with a cubic optical lattice.
There the $\Mott$ state is a candidate for quantum information processing \cite{porto2003quantum} tasks and quantum simulation of spin systems \cite{hild2014far}, but experimental protocols for the initial lattice loading leaves the system in the $\SF$ state. 
Within this physical implementation of the Bose-Hubbard model, the occupation number $n_i$ is associated with the lowest band Wannier function maximally localized on site $i$, see Appendix~\ref{app:modelling} for a more detailed discussion. 
The characteristic energies $J_x(v_x)$ and $U(v_x,v_y,v_z)$ are then implicitly related to the lattice trapping depths $v_x,v_y,v_z$ as denoted and for fixed $v_y, v_z$ this translates into a functional dependence $v_x(U/J_x)$.
These monotonic mappings are calculated numerically for a set of experimentally relevant lattice parameters in Appendix~\ref{app:modelling}
[$J_x$ is exponentially decreasing with $v_x$ and 
$v_x$ is exponentially increasing with $U/J_x$]. 

Matrix product states turn out to be an effective description one-dimensional many-body systems. 
The price paid for such extended simulatory treatments is a significant increase in analytical and numerical code logistics. 
See Appendix~\ref{app:mps} for a more detailed discussion of matrix product states. 

\subsection{Exact gradient optimization}
\label{sec:exactgradient}
Here we briefly review derivative-based optimal control and main results of Ref.~\cite{jensen2021approximate} in the context of the present work.  \\

In quantum optimal control we seek to dynamically steer some quantum mechanical process in a controlled way such as to maximize a desired physical yield. For unitary evolution, any such task can be encoded as a minimization over an appropriate cost functional $J[\U(T;0)]$ where 
\begin{align}
\U(T;0) = \mathcal{T} \exp\bigg(-i \int_{0}^{T} \H(t) \mathrm{d}t\bigg), \label{eq:UOrdered}
\end{align}
is the time evolution operator in units where $\hbar = 1$ from time $t=0\rightarrow T$, $\mathcal T $ denotes time ordering, 
and $\op H$ is the system Hamiltonian carrying some generic time dependence.
The Hamiltonian can be generically decomposed as 
\begin{align}
\H = \H(t,u(t)) = \Hd(t) + \Hc(t,u(t)), \label{eq:generalH}
\end{align}
where $\Hd$ is the uncontrollable drift Hamiltonian and $\Hc$ is the control Hamiltonian parametrized by the {control}  $u(t)$
which allows manipulation of the unitary time evolution. The extension to more than one control parameter is discussed in Ref.~\cite{jensen2021approximate}. 
Discretizing time on a regular grid of length $\nt$ spaced by $\dt$ we obtain
\begin{subequations}
	\label{eq:timeevolution} 
	\begin{align}
	&t \in [t_1,t_2,\dots,t_{\nt}] = [0, \dt, \dots, T], \quad  t_j = (j-1)\dt, \\
	&\U(T;0) \approx \prod_{j=1}^{\nt-1} \U_j = \U_{\nt-1}\dots \U_2\U_1, 
	\end{align}  
\end{subequations}
with time indices denoted as subscripts and $\U_n$ is the propagator across the time interval $[t_n, t_{n+1}] = [t_n, t_n+\delta t]$.
The propagators $\U_n$ depend on the discretized control vector  $\vec u= (u_1,\dots,u_{\nt})$ 
in some manner depending on the chosen discretization scheme. 
We can then numerically minimize $J(\vec u)=J(\U_{\nt-1}\dots \U_2\U_1)$ through iterative local updates of the control vector
\begin{align}
&\vec u^{(k+1)} = \vec u^{(k)} + \alpha^{(k)} \vec p^{(k)}, \label{eq:localupdate}
\end{align}
such that $J(\vec u^{(k+1)}) \leq J(\vec u^{(k)})$. 
The step size $\alpha^{(k)} > 0$ at iteration $k$ is found by line searching and there are several choices for the search direction $\vec p^{(k)}$ which all depend on the local gradient $\vec{\nabla} J(\vec u^{(k)})$. 
This includes e.g. the steepest descent, \bfgs, and Newton direction \cite{nocedal2006numerical}.

A local minimizer $\vec u^*$ of $J$ is called an optimal control. 
The cost typically contains several contributions, $J=\sum_i J_i$, where each term encodes a desired feature that prospective optimal controls strives to fulfill.
This always includes a term explicitly related to the quantum dynamics, $J\st{dyn}(\U_{\nt-1}\dots \U_1)$, and potentially various control constraints, $J\st{con}(\vec u)$, that do not (see Appendix~\ref{app:opt}). 
The total gradient is then given by the sum of the individual gradients, $\vec \nabla J=\sum_i \vec \nabla J_i$. 
The functional form of the cost $J\st{dyn}$ depends on the particular unitary task \cite{schirmer2011efficient}, yet the associated gradient elements can always be reduced to evaluating expressions on the form
\begin{align}
\frac{\partial J\st{dyn}}{\partial u_n}  \sim \frac{\partial}{\partial u_n} \left( \U_{\nt-1}\dots \U_2 \U_1\right). \label{eq:chainrule}
\end{align}
Evaluating the right hand side of \eqref{eq:chainrule} constitutes the most numerically expensive computation in this optimization paradigm and is what leads to the typical computational bottlenecks mentioned in the introduction. 
Once calculated, however, it is straightforward to assemble the gradient for any $J[\U]$. 
A main result of Ref.~\cite{jensen2021approximate} is that Eq.~\eqref{eq:chainrule} can be efficiently evaluated by
\begin{subequations}
\label{eqs:maindynamics}
\begin{align}
\frac{\partial}{\partial u_n} \left( \prod_{j=1}^{\nt-1} \U_j \right)  = \left(  \prod_{j=n}^{\nt-1} \U_j  \right) \left(-i\delta t \frac{\partial H_n^c}{\partial u_n}\right) \left(\prod_{j=1}^{n-1} \U_j \right), \label{eq:derivativeU}
\end{align} 
given that the system is described in a basis where $\H^c_n=\H^c(t_n,u_n)$ is diagonal and simultaneously employing a specific \footnote{The perhaps more familiar Trotterization $\exp\bigg({-i\H_{n}^c \dt/2}\bigg)\exp\bigg({-i\H_{n}^d \dt}\bigg) \exp\bigg({-i\H_{n}^c \dt/2}\bigg) $ leads to a similar but distinct expression, see Ref.~\cite{jensen2021approximate}.} Trotterized propagator,
\begin{align}
\U_n &= \U_n\ut{ST}  \equiv \U_{n+1}^{c/2} \U_n^d \U_{n}^{c/2} 
\end{align}%
where we defined the control and drift propagators 
\begin{align}
\U_n^{c/2} &\equiv \exp\bigg({-i\H_{n}^c \dt/2}\bigg),  \quad\U^d_n \equiv 	\exp\bigg({-i \Hdbar\dt}\bigg),
\end{align}
\end{subequations}
and $\Hdbar = \frac{1}{2}(\Hd_{n+1}+\Hd_{n})$ where $\H^d_n = \H^d(t_n)$. 
  \medskip

The $\SF\rightarrow \Mott$ state transfer described in Sec.~\ref{sec:bhmps} is conveniently encoded as a maximization of the fidelity or minimization of an associated $J\st{dyn}$ cost, respectively given by \cite{jensen2021approximate}
\begin{subequations}
\begin{align}
F& =|\braket{\psitgt| \psi(T)}|^2 = |\braket{\psitgt| \U(T;0) |\psio}|^2 \label{eq:F}, \\
J_F &= \frac{1}{2}\left(1 - F\right),  \label{eq:J} 
\end{align}
\end{subequations}
where $\ket{\psitgt} = {\Mott}$ is the target state, $\ket{\psio}={\SF}$ is the initial state, and $\ket{\psi_{\nt}} = \ket{\psi(T)} = \U(T;0) \ket{\psio}$ is the time-evolved state at final time $T$. 
Using Eqs.~\eqref{eqs:maindynamics} the analytically exact gradient elements \footnote{An additional factor $1/2$ is present for the end points $n=1,\nt$.} of the fidelity cost can then be written as 
\begin{subequations}
	\label{eq:ST}
\begin{align}
\frac{\partial J_\F}{\partial u_n} &= \Re \left(i \overlap^*\Braket{\chi_{n} | \frac{\partial\H_n^{c}}{\partial u_n}| \psi_{n}  } \right)\dt, \label{eq:exactGradientST} 
\end{align} 
where $\ket{\psi_{n+1}} = \U_n\ket{\psi_n}$ is the forward propagated initial state $\ket{\psi_1}= \ket{\psio}$ while 
$\ket{\chi_{n}} = \U_{n}^\dagger \ket{\chi_{n+1}}$ is the backward propagated target state $\ket{\chi_{\nt}} = \ket{\psitgt}$ with the transfer amplitude $\overlap =\Braket{\chi_{\nt}|\psi_{\nt}}$. 
The exact gradient computation is only dominated by the time it takes to solve the dynamics --- once forwards and once backwards. Any optimal control algorithm, including derivative-free ones, is in each iteration bounded from below by this propagation time. This gradient thus saturates this bound up to a small constant scaling (as opposed to other known exact gradients). 

It is noted that a variational treatment of the cost functional prior to temporal discretization leads to a gradient expression that is superfically similar to Eq.~\eqref{eq:exactGradientST}, but such an approach assumes an infinitesimal time step. 
Time steps are however always finite in numerical practice.  This subtlety means that the details of the chosen propagator are important and that the ``well-known'' variationally obtained gradient is only a first-order approximation. 
It is then only under the conditions surrounding Eqs.~\eqref{eqs:maindynamics} that the corrections vanish and Eq.~\eqref{eq:exactGradientST} is exact which
by happenstance coincides with the approximate variational calculation. 
See Ref.~\cite{jensen2021approximate} for an expanded discussion. \\


%

The control may also be decomposed on a set of parametrized functions $\{f_l(t;\vec \theta)\}_{l=1}^L$ 
\begin{align}
u(t; \vec \theta) = f_0(t) + \sum_{l=1}^{L} f_l(t; \vec \theta), \label{eq:crab}
\end{align}
where $f_0(t)$ is a fixed reference function and $\vec \theta \in \mathbb{R}^M$ are now the optimization parameters. 
This is widely referred to 
\footnote{The name implies that the set of functions becomes a complete basis as $L\rightarrow \infty$, but in numerical practice this is not a necessary requirement.} 
as the ``chopped random basis'' (\crab) technique \cite{caneva2011chopped,muller2021one}, 
and a typical choice of functional basis consists of ``randomized'' Fourier components. 
By choosing an appropriate set of functions the search space can be both significantly reduced and focused on certain ``realistic'' control shapes depending on the application. 
Exact gradient optimization in a \crab~ parametrization is also admitted by Eqs.~\eqref{eqs:maindynamics}
through the chain rule, 
\begin{align}
\frac{\partial J(\vec u(\vec \theta))}{\partial \theta_i} &= \sum_{n=1}^{\nt} \frac{\partial J}{\partial u_n}  \frac{\partial u_n}{\partial \theta_i} = \sum_{j=1}^{\nt} \sum_{l=1}^{L} \frac{\partial J}{\partial u_n} \frac{\partial f_l(t_n;\vec \theta )}{\partial \theta_i}.  \label{eq:group}
\end{align}
\end{subequations}
The gradient element with respect to the parameter $(\vec \theta)_i=\theta_i$ therefore still depends on the ``nonparametrized'' gradient elements $\partial J/\partial u_n$ described above with negligible further computational overhead. 
\\

In applying the exact gradient methodology here, it is convenient to rescale Eq.~\eqref{eq:H} at time $t_n$ by the instantaneous tunneling rate, 
\begin{subequations}
\begin{align}
\H_n &= \frac{\H\st{SI}(t_n)}{J_x (t_n)}  =  \sum_{i=1}^{N_s-1} \h^{J_x}_{[i,i+1]} + \frac{1}{2}\left(\frac{U(t_n)}{J_x(t_n)}\right)\sum_{i=1}^{N_s} \h^U_{[i]} \label{eq:bhcontroldrift}. 
\end{align} 
We identify the drift and control Hamiltonian as respectively
\begin{align}
\H^d = \sum_{i=1}^{N_s-1} \h^{J_x}_{[i,i+1]}, \qquad \H_n^c = \frac{u_n}{2}\sum_{i=1}^{N_s} \h^U_{[i]},
\end{align}
where $u_n = U(t_n) / J_x(t_n)$ is the control parameter. 
The control Hamiltonian is thus diagonal, as required for the validity of Eqs.~\eqref{eq:derivativeU} and \eqref{eq:exactGradientST}, 
when represented in the site-occupation basis. 
As an additional numerical benefit, the drift Hamiltonian is time-independent, $\Hdbar = \Hd$, and so is the control Hamiltonian derivative needed in Eq.~\eqref{eq:exactGradientST},
\begin{align}
\frac{\partial \H_n^{c}}{\partial u_n}  &= \frac{1}{2}\sum_{l=1}^{N_s} \h^U_{[i]}.  \label{eq:bhcontrolderiiv}
\end{align}
This structure of Eq.~\eqref{eq:bhcontroldrift} can be exploited to significantly accelerate the 
time evolution of the matrix product state as described in Appendix~\ref{app:mps}.  \\
\end{subequations}

The control vector $\vec u =(\dots, u_n, \dots )  = (\dots, U(t_n) / J_x(t_n), \dots )$ is in our simulations independent of the physical implementation of the Bose-Hubbard model. 
It can be translated into a corresponding time sequence of laboratory parameters such as, e.g., optical lattice depth $\vec v_x(\vec u) = (\dots, v_x(u_n), \dots)$ by specifying the relationship between $v_x$ and $U/J_x$ as discussed in the Appendix. 
The value $v_x(u_n)$ must be applied for a duration proportional to $\hbar/J_x(u_n)$, and the total duration measured in SI units therefore depends on the control $T\st{SI}(\vec u)$.  
The translated $\vec v_x$ is therefore qualitatively speaking a scaled version of the corresponding $\vec u$.  

In the following we denote durations by the format $T=T\st{sim}\,(T\st{SI})$ where $T=T\st{sim}$ is the process duration entering in Eqs.~\eqref{eq:timeevolution}. 
The presented SI numbers are based on the set of experimental lattice parameters described in Ref.~\cite{van2016optimal}, see Appendix~\ref{app:modelling}.   

\begin{figure}[b]
	\includegraphics[width=\linewidth]{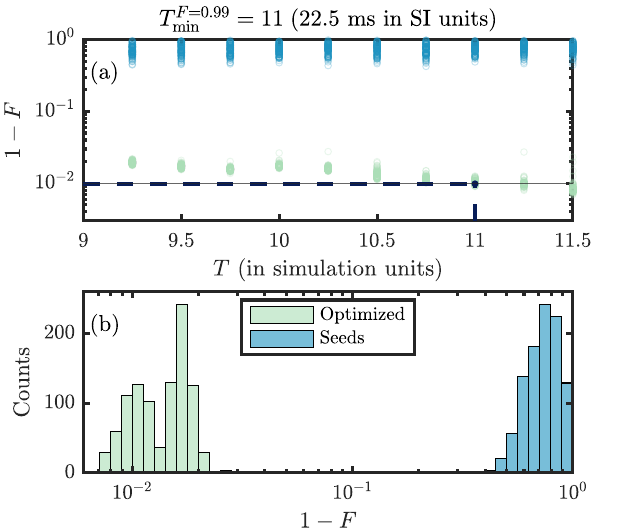}
	\vspace*{.2cm}
	
	\includegraphics[width=\linewidth]{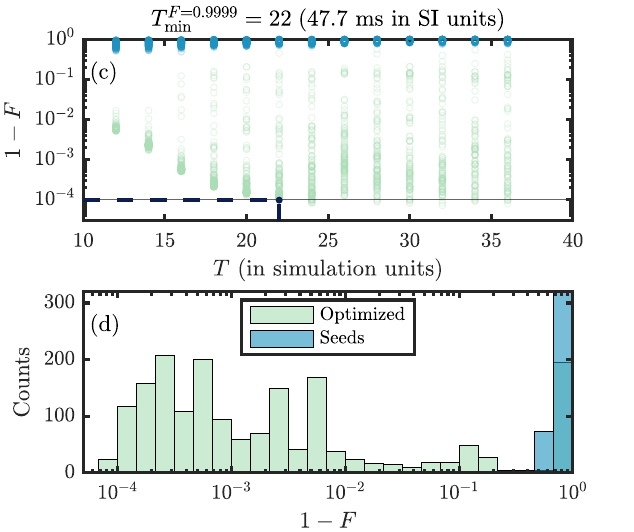}
	\caption{ Optimization results for $N_s = N_p = 20$ showing infidelity $1-F$ (lower is better) for each seed before (blue) and after (green) optimization. 
		(a): Each dot represents a control and the translucency informs about the distribution density. 
		In this batch we find $\tqsl[0.99] =  11\, (22.5\,\si{ms})$ indicated with the dashed line.	
		(b): Histogram (left is better) prior to and after optimization with the same color scheme.
		(c)--(d): As (a) and (b), but at higher durations and finding $\tqsl[0.9999] =  22\, (47.7\,\si{ms})$. 
	}
	\label{fig:FT99}
\end{figure}

\subsection{Results}
\label{sec:results}
We now present exact gradient optimization results for the $\SF \rightarrow \Mott$ many-body transfer at unit filling $N_p = N_s=20$ using matrix product states. 
A numerical exact diagonalization treatment \cite{zhang2010exact} is prohibitively resource intensive at these numbers as the Hilbert space has dimensionality $D_\mathcal{H} \sim 10^{11}$ and requires $\sim\!1\,\si{TB}$ memory just to store a single generic state. 
The initial starting points (seeds) for the optimization were generated by adding a sum of randomized Fourier components to an adiabatically inspired reference control, i.e. an exponential function that slowly crosses the critical point before ramping up.
After generating a seed it is optimized using Eq.~\eqref{eq:exactGradientST} --- it is not bound by a parametrization and each $u_n = u(t_n)$ is independently adjustable. 
Appendix~\ref{app:opt} gives an overview of further optimization details.

\begin{figure}[h!]
	\vspace{0.2cm}
	\includegraphics[]{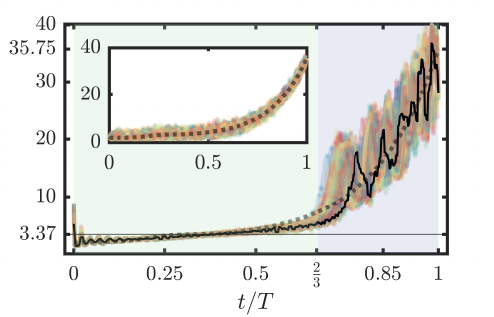}
	\caption{Sample duration-normalized control ramps (multi-colored faded lines) $u(t/T)$ for solutions with $F\geq 0.99$. The corresponding seeds are shown in the inset, the black dotted line indicates the adiabatically inspired reference control, the horizontal line denotes the critical point value for the phase transition (see Appendix), and the black solid line highlights the optimal control with duration $\tqsl[0.99] = 11 \,(22.5\,\si{ms})$. The optimized controls are characterized by two distinct segments denoted by the shaded areas.}
	\label{fig:controls}
	\vspace{0.6cm}
	\centering
	\includegraphics[]{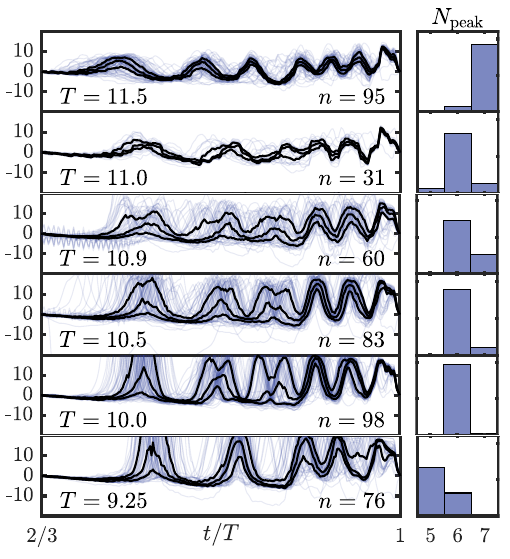} 

	\caption
	{
		Distinct optimal control strategies characterized by a different number of peaks/bangs are active at different $T$ (rows). The left panels show $n$ found high-fidelity controls (translucent lines) and their (25,50,75)\% quantiles (black lines). The right panels show the relative distribution of the $n$ solutions into the three identified strategies at these $T$.  
	}
	\vspace{0.5cm}
	\label{mps:fig:transformedcontrols}
\end{figure}

Figure~\ref{fig:FT99} shows the empirically achieved fidelity through numerical optimization as a function of process duration $T$. 
From these we find minimal duration estimates $\tqsl[0.99] = 11\,(22.5\,\si{ms})$ and $\tqsl[0.9999] = 22\,(47.7\,\si{ms})$. 
The highest fidelities as a function of $T$ are found to approximately follow an exponential law. 
The effect of the optimization is seen to be significant as solutions are typically improved by several orders of magnitude. 
Sample optimization trajectories are shown in Appendix~\ref{app:opt}. 

This successful application of exact-gradient optimal control techniques in the many-body limit constitutes the first main result of this paper. 
\\

Figure~\ref{fig:controls} shows a subset of the optimized control ramps from Fig.~\ref{fig:FT99}(a) and we observe two control segments with very distinct behaviors. 
\pgfmathsetmacro{\theitemindent}{-0.25}
\begin{itemize}[itemindent=\theitemindent em,align=left,  leftmargin=*]
	\item[\textbf{Segment 1 -- ``linear'':}  $\mathbf{t/T \in [0, \frac{2}{3}]}$.] ~\newline
	With nearly vanishing variance and following an initial perturbation, this predominantly linear segment ``slowly'' crosses the critical point with high-frequency/low-amplitude oscillations. 
	
	\item[\textbf{Segment 2 -- ``bang-bang'':} $\mathbf{t/T \in [\frac{2}{3}, 1]}$. ] ~\newline
	With large variance, this segment ramps to the final control value and consists of several low-frequency/high-amplitude oscillations.	
	These oscillations turn out to be smoothed approximations to bang-bang structure. 
\end{itemize}
Focusing on segment 2,  subtracting the linear contribution there, and resolving the optimized solutions according to their duration $T$,  we obtain a clearer picture in Fig.~\ref{mps:fig:transformedcontrols}. 
Namely, there are several optimal control \textit{strategies} \cite{jensen2020crowdsourcing} that are individually characterized by an integer number of peaks/bangs $N\st{peak}=5,6,7$. 
The best and most prominent strategy depends on $T$ where higher $T$ leads to a higher optimal number of peaks.
This trend is verified to continue outside of this particular range of durations where different $N\st{peak}$ are prevalent. 
At lower $T$ the peaks are increasingly bang-bang-like while at higher $T$ the peak shapes are smoother and less extreme \footnote{This may be partially due to the regularization discussed in Appendix~\ref{app:opt} not being duration-normalized.}.  
Since the effect on the dynamics is largely the same, however, we will for simplicity refer to these peaks collectively as bangs although they may not always meet the strictest definition. 
Translating $u(t) = U(t) / J_x(t)$ into e.g. optical lattice depths $v_x(t)$ scales the shape but leaves the overall structure intact --- the peaks then appear to have roughly the same width $\approx1.3\, \si{ms}$ and center distance $\approx2-2.5\, \si{ms}$ independently of $T$.
The recoil energy time scale is roughly $h/E_R \approx 0.5\,\si{ms}$ for comparison.

The discovery of this family of optimal solutions for the $\SF\rightarrow\Mott$ transfer constitutes the second main result of this paper. 
The existence of other types of optimal solution shapes cannot be excluded and we leave these types of searches for future work.
\\

To understand the physical mechanisms at play in the problem, we inspect the quantum dynamics induced by these fidelity-optimized controls at different durations in Fig.~\ref{fig:observables}. 
From $\Braket{\op n_i}_{\psi(t)} = \Braket{\psi(t) |\op n_i |\psi(t)}$ in Fig.~\ref{fig:observables} we find that a light-cone-like homogenization of site population takes place when crossing 
the critical point in segment 1. The population is otherwise initially concentrated in the bulk (roughly sites 4-17) due to finite edge effects.
This process is clearly limited by the propagation velocity $\partial \Braket{\op n_i}_{\psi(t)}/\partial t$ and thus also influences the minimal duration $\tqsl[]$ for high-fidelity transfers. 
The tardiness hereof is somewhat expected because we are crossing a phase transition and 
\noindent the adiabatic time scales approaches infinity as discussed earlier.
Even upon reaching a largely homogenized site population after segment 1, however, the fidelity remains very small. 
The following bang-bang process in segment 2 corresponds to alternating between tunneling events and site-locking --- during the latter, individual $c$-phases are imprinted on the individual Fock components 
\begin{align}
\!\U_n &\ket{n_1,\dots, n_{N_s}}  \approx \notag\\ 
&\ket{n_1,\dots,n_{\nsites}} \cdot \exp\left({-\frac{i u\dt}{2}\sum_{i=1}^{\nsites}n_i(n_i-1)}\right)
, \label{mps:eq:phaseacq}
\end{align}
since tunneling is approximately negligible in the deep lattice limit, $u\lesssim  u\ut{max}$, see Appendix~\ref{app:modelling}.
This leads to a nontrivial interplay ($F(t)$ is nonmonotonic) between the canonically conjugate population and phase variables that evidently
produces the correct interferences finally leading to the target $\Mott$ state. 
\\

It is interesting to note that a recent work \cite{brady2021optimal} argues that optimal controls should exhibit a ``bang-smooth-bang'' structure for a certain class of problems.
Said problems are equivalent to special cases of the state transfer formulation 
where the total Hamiltonian, $\H(u(t)) = u(t)\H_1 + (1-u(t)) \H_2$, is a linear interpolation with $0\leq u(t) \leq 1$ between Hamiltonians $\H_1$ and $\H_2$, and $\ket{\psio} = \ket{\mathrm{GS}; \, u=1}$ and $\ket{\psitgt} = \ket{\mathrm{GS};\,  u=0}$ are the ground states for $\H_1$ and $\H_2$. 
The problem studied here is approximately on this form and it appears plausible that our optimized controls are related to this bang-smooth-bang prediction.

 \subsection{Comparison to previous open-loop efforts}
 We now discuss our two main results in relation to previous open-loop efforts \cite{doria2011optimal,van2016optimal} of optimizing the $\SF \rightarrow \Mott$ transition. 
 These explicitly assume an optical lattice implementation and employed derivative-free Nelder-Mead methodologies to optimize the expansion coefficients of a Fourier component \crab. 
 The respective optimization results and methodologies are discussed in turn.  \\

\clearpage

\makeatletter\onecolumngrid@push\makeatother
\begin{figure*}
	{
		\includegraphics[trim={0cm 0.9cm 0cm 0cm},clip]{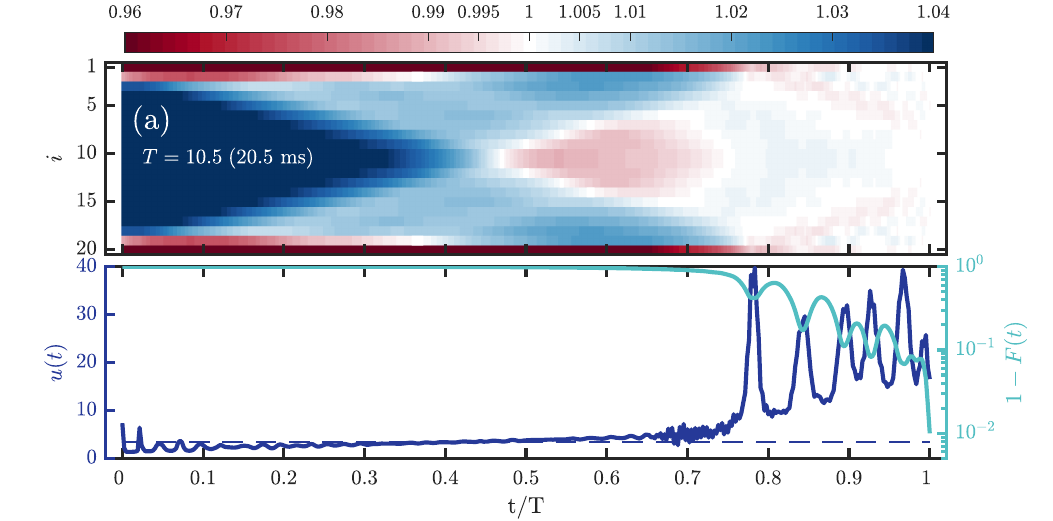}
	}
	{
		\includegraphics[trim={0cm 0.9cm 0cm 0.95cm},clip]{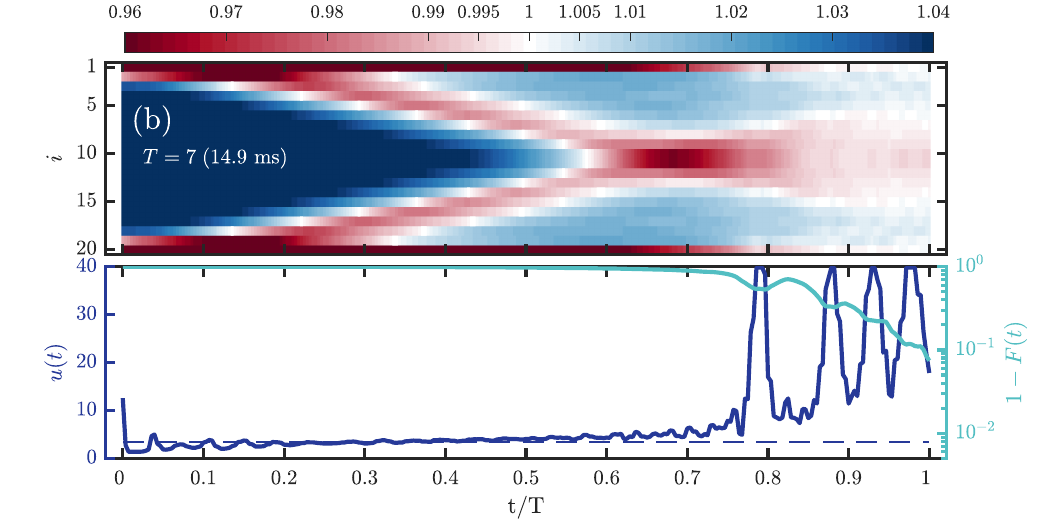}
	}
	{
		\includegraphics[trim={0cm 0cm 0cm 0.95cm},clip]{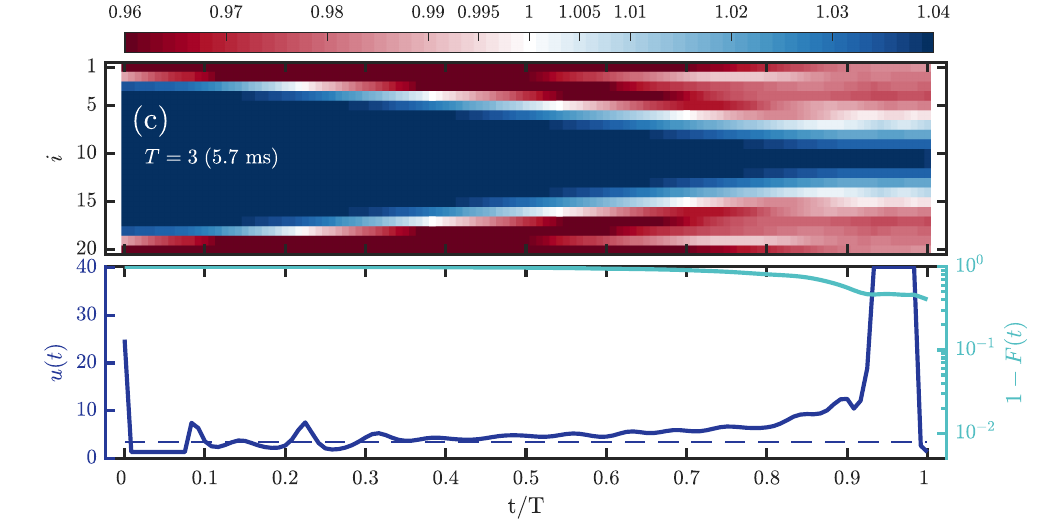}
	}
	\caption{
		(a)--(c): Time evolution along sample optimized controls at three different durations. Each upper panel shows the site occupations $\Braket{\op n_i}_{\psi(t)}$ (sensitive near unit occupancy and note $\Braket{\op n_1}_{\psi(0)} = 0.67$) while each lower panel shows the corresponding control (solid purple line, left axis) and critical value $(U/J)\st{crit}$ (dashed purple line, left axis) and infidelity (solid teal line, right axis).
		The $t$ axis is slightly extended to better display the behavior at $t/T=0,1$. 
	}
	\label{fig:observables}
\end{figure*}
\clearpage
\makeatletter\onecolumngrid@pop\makeatother

Our first result was to show that exact gradients are now a viable option in the many-body limit, which was previously considered potentially infeasible \cite{doria2011optimal}. 
Direct quantitative comparisons to Refs.~\cite{doria2011optimal,van2016optimal} in terms of fidelity and process durations are obstructed by the differences in the exact problem formulations summarized in Table~\ref{tab:comparison}. 
Nevertheless, the conservative quantitative inferences discussed below indicate that the new optimization technique is indeed promising. 
Strong comparative conclusions between optimization methodologies can only be drawn if they are applied in numerical environments where everything else is equal. 
We consider such extended numerical comparisons outside the scope of this first demonstration of exact gradients in the many-body limit. 

The optimized observables reported in Refs.~\cite{doria2011optimal,van2016optimal} are not the $\Mott$ fidelity. 
Instead they are related only to the site population statistics and are thus insensitive to the relative phases between Fock state components. 
They are thus correlated with the fidelity but not in one-to-one correspondence. 

\begin{figure}
	\centering
	\includegraphics[trim={0.3cm, 0.3cm, 0.4cm, 0cm},clip]{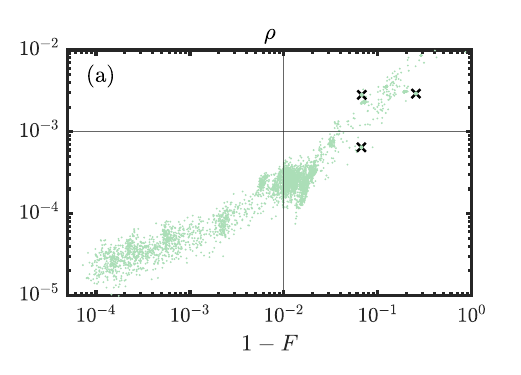}
	\includegraphics[trim={0.0cm, 0.0cm, 0.0cm, 0cm},clip]{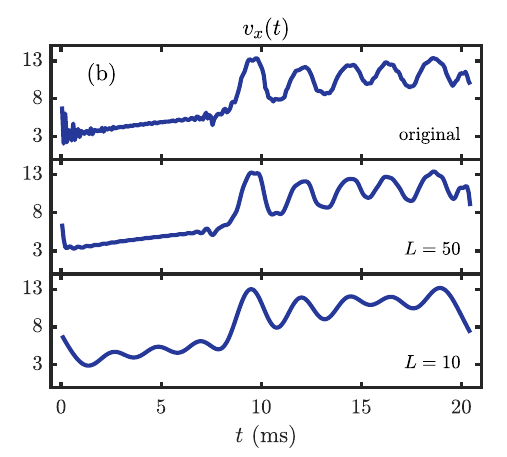}
	\caption{
		(a): Density of defects Eq.~\eqref{eq:rho} for all the fidelity-optimized solutions (green dots). The horizontal line denotes $\rho_c = 10^{-3}$ from Ref.~\cite{doria2011optimal} and the vertical line denotes $F=0.99$. The black crosses highlight solution pairs that have nearly the same $\rho$ but very different $F$ and vice versa. 
		(b): Optical lattice depths $v_x(t)$ mapped from $u(t)$ in Fig.~\ref{fig:observables}(a) and the corresponding signal when 
		removing Fourier components with a frequency larger than $\nu_L = 2\pi L /T$. 
		The $y$ axis is in units of recoil energies using lattice parameters from Ref.~\cite{van2016optimal}. 
	} 
	\label{fig:comparison}
\end{figure}

Reference~\cite{doria2011optimal} reports on the final density of defects,
\begin{align}
\rho &= \frac{1}{N_s}\sum_{i=1}^{N_s}| \braket{\op n_i}_{\psi(T)} -1|. \label{eq:rho}
\end{align}
Vanishing $\rho$ values \footnote{Our chosen target state has a defect density of $\rho \approx 2.8 \times 10^{-6}$.}  are indicative of a $\Mott$ state and optimization of a solution in Ref.~\cite{doria2011optimal} was halted if it achieved $\rho \leq \rho_c = 10^{-3}$.
The density of defects should not be affected by the differences in Table~\ref{tab:comparison}, and this allows a point of comparison when evaluating $\rho$ for all our fidelity-optimized solutions as shown in Fig.~\ref{fig:comparison}(a).
We are able to verify that a fixed value of $\rho$ can display a large variance in $F$ and vice versa. 
For example, $\rho\approx3\times 10^{-3} \geq \rho_c$ can correspond to both $F\approx 0.75$ and $F\approx0.93$
while $F\approx0.93$ can also correspond to $\rho=6\times 10^{-4} \leq \rho_c$.
Thus $\rho_c$ is not an ideal stopping condition for the $\Mott$ state in terms of fidelity and it is not possible to know what the fidelity distribution in Ref.~\cite{doria2011optimal} would look like. 
However, all our solutions with $F\geq 0.99$ have less than $\rho \approx 4\times 10^{-4}$ and solutions with $F\approx 0.9999$ are distributed near $\rho \approx 2 \times 10^{-5}$. 

Our numerical calculation of the relationship $v_x(U/J_x)$ described in Appendix~\ref{app:modelling} --- which is used to calculate SI times for a given set of lattice parameters --- yields values that are consistent with those stated in Ref.~\cite{van2016optimal}.
When we instead use the tighter lattice parameters in Ref.~\cite{doria2011optimal} the SI durations are reduced by a roughly factor of 2--3. 
This places our process durations on the order of $\si{ms}$ which is similar to the $3.09\,\si{ms}$ used in Ref.~\cite{doria2011optimal}.
There is, however, not a basis for direct comparison between these since we observe a discrepancy in $v_x(U/J_x)$ at the values reported in \cite{doria2011optimal}.
(The variance between values of $F$ and $\rho$ above suggests that $F\geq 0.99$ requires more than $3.09\,\si{ms}$ in the geometry of Ref.~\cite{doria2011optimal}.) \\

The work in Ref.~\cite{van2016optimal} assumes $N_p = 16, N_s = 32$, and includes a harmonic potential $\Omega \times \sum_{i=1}^{N_s} (i - i_0)^2 \op n_i$
where $i_0 = (N_s - 1)/2$ and $\Omega=2.4 \times 10^{-3} E_R$ which is weak compared to the on-site interaction $U$ at all lattice depths (see Fig.~\ref{fig:UJ}). 
Note that Ref.~\cite{doria2011optimal} shows optimization results both with and without a similarly weak harmonic potential and in that instance there does not seem to be a significant difference between the quantitative outcomes.

The observable reported and minimized in Ref.~\cite{van2016optimal} is the rescaled average in the center of the trap
\begin{align}
\eta &= \frac{1}{8} \sum_{i=N_s/2 - 3}^{N_s/2+4} \frac{ \Delta \op n_i(T)}{ \Delta \op n_i(0)}, \label{eq:eta}
\end{align}
where $\Delta \op n_i(t) = \braket{\op n_i^2}_{\psi(t)} -\braket{\op n_i}^2_{\psi(t)}$. 
The smallest value reported is $\eta \approx 1 \times 10^{-1}$ at roughly $15\,\si{ms}$. 
Similarly to $\rho$ we verify in our data that a fixed value of $\eta$ (where we extend the sum to all lattice sites) can exhibit a large variances in $F$.
We observe $\eta = 1 \times 10^{-1}$ at $F=0.75$ while all fidelities $F\geq 0.99$ have $\eta \leq  1.4 \times 10^{-2}
$ and at most reached $\eta \approx 1.2 \times 10^{-2}$. 
Since the harmonic potential is quite weak and that the reported $\eta$ more leniently covers only 8 sites, these numbers suggest that it is quite unlikely that the corresponding states in Ref.~\cite{van2016optimal} are close to $F=0.99$ with respect to the ground state. 
In terms of process durations the $F=0.75$ control in Fig.~\ref{fig:observables}(c) attains $\eta \approx 1 \times 10^{-1}$ at about a third of the duration in Ref.~\cite{van2016optimal}. \\

These assessments indicate that our exact gradient methodology is able to achieve results that are competitive with the methodologies employed by Refs.~\cite{doria2011optimal,van2016optimal} in terms of solution quality and transfer times. 
\\

Our second main result was that the $\SF \rightarrow \Mott$ transition has a family of solutions consisting of distinct segments  
consisting of a predominantly linear sweep followed by bangs. This overall control structure is preserved when mapped to optical lattice depths as seen in Fig.~\ref{fig:comparison}(b). 
From a qualitative point of view, the optimized controls reported in Refs.~\cite{doria2011optimal,van2016optimal} are in contrast very smooth and are generally similar to each other. 
A more quantitative statement is that our obtained optimized controls have a higher information-theoretic \textit{control complexity} as initially introduced in Ref.~\cite{caneva2014complexity} where an operational definition is given by the number of Fourier components needed to solve the problem to a given fidelity threshold (these and associated notions were shortly after treated in more generality in Ref.~\cite{lloyd2014information}). 
A partial reason for the observed disparity can be understood as follows. 

As mentioned, Refs.~\cite{doria2011optimal,van2016optimal} are based on the \crab~ technique in Eq.~\eqref{eq:crab}
with derivative-free Nelder-Mead optimization.
They utilize Fourier components centered on low-lying harmonic frequencies, e.g. $\nu_l = 2\pi l(1+r_l)/T$ where $r_l \in[0:1]$ are randomized frequency offsets \cite{doria2011optimal}. 
That is, the basis functions are on the form $f_l(t, \vec \theta) \propto \theta_{l}^{\sin} \sin\nu_l t + \theta_{l}^{\cos} \cos\nu_l t$ 
where the control $\vec \theta = (\vec \theta^{\sin}, \vec \theta^{\cos}) = (\theta_1^{\sin}, \dots, \theta_L^{\sin},  \theta_1^{\cos}, \dots, \theta_L^{\cos})$ represents the $M=2L$ expansion coefficients.
The number of components used in Refs.~\cite{doria2011optimal,van2016optimal} is stated as small but is not explicitly specified.  
A low-dimensional Fourier decomposition effectively means that the control signal is bandwidth/rate of change limited. 
Our presented optmizations instead incorporate rate of change limitations through a cost functional, see the Appendix. 
The restrictions imposed in this way are somewhat more difficult to characterize, but it is observed that the maximal rate of change due to the approximate bangs in e.g. Fig.~\ref{fig:observables} is similar to the maximal rate of change near $t=T$ in Refs.~\cite{doria2011optimal,van2016optimal}. 

Regardless of bandwidth considerations, the restriction to a small $M$ in Refs.~\cite{doria2011optimal,van2016optimal} is necessary because the heuristic Nelder-Mead methodology does not generally perform well as $M$ becomes large \cite{nocedal2006numerical}. 
Reference~\cite{sorensen2018quantum} reports, for example, on a similar parametrization for a different control problem, and the statistically best performance for Nelder-Mead was limited to $F\approx 0.9$ at only $M\approx 20$ (the performance deteriorated for higher $M$). 
Results in the same parametrization obtained by gradient-based means showed statistically, however, that roughly a factor 2-3 more components were at minimum required to resolve $F=0.99$ controls. 
A similar number of  components was needed in Ref.~\cite{sorensen2020optimization}.
Assuming that $M\approx 20$ is a reasonable general guideline for Nelder-Mead's maximum basis size leaves only the lowest $L=M/2\approx 10$ harmonic frequencies for the expansion in Refs.~\cite{doria2011optimal,van2016optimal}. 
The spectral cutoffs \footnote{The spectral cutoffs were achieved by Fourier transforming the signal, zeroing components with frequency larger than $\nu_L$, and transforming the signal back to the time domain.} in Fig.~\ref{fig:comparison}(b) show that this is insufficient to represent the control structure yielding $F=0.99$ from Sec.~\ref{sec:results}, and roughly a factor five more frequency components are needed.


The initial practicality of restricting the search space to a limited set of basis functions such as Fourier components may therefore not be ideal because there is no guarantee that the ultimately best solutions are captured by the parametrization. 
A typical extension known as dressed-\crab~(\dcrab) \cite{rach2015dressing} attempts to remedy this phenomenon by drawing a new set of basis functions, e.g. by sampling new random frequency offsets for the Fourier components. 
This can be expected to work reasonably well only if the overall parametric structure can capture the solutions. 
In this case, however, a significant increase in the spectral content is needed to produce the structure in Sec.~\ref{sec:results}. 
This cannot be accounted for by small perturbations around a limited number of low harmonic frequencies.
The Fourier \crab~ can only attain the required control complexity by increasing the number of components $L$, but this is at odds with Nelder-Mead requiring a small number of optimization parameters. 
A derivative-free Nelder-Mead with a ``standard'' Fourier \crab~is thus likely prohibited in identifying the high-fidelity optimal controls in Sec.~\ref{sec:results}. 

The discussion above highlights the potential drawbacks of preemptively choosing a parametrization. 
It does not imply that parametrization are always detrimental.  
For a given parametrization, however, we expect the gradient-based update rule in Eq.~\eqref{eq:group} to statistically yield better results than a gradient-free update rule in the ideal open-loop context \cite{sorensen2018quantum} (the opposite may be true in the closed-loop context and this leads into their potential unification discussed in Sec.~\ref{sec:unified}). 
It is also straightforward to calculate exact gradients with and without parametrizations for $\rho$ and $\eta$ observables in Eqs.~\eqref{eq:rho}-\eqref{eq:eta}. 
Since these observables are much more lenient than fidelity it is expected that $\rho=10^{-3}$ and $\eta=10^{-1}$ solutions can be found at significantly reduced process durations, but that these will not correspond to high-fidelity solutions. 

Despite the potential issues encountered by Nelder-Mead, it is quite clear that the optimized solutions in Refs.~\cite{doria2011optimal,van2016optimal} improve significantly on both the observable metrics and transfer times compared to adiabatic solutions.  
This fact can be interpreted based on the present analyses as follows. 
Recall that the control strategies were characterized by a predominantly linear sweep followed by a number of bangs where a lower number corresponds to lower durations and fidelity. 
Fewer bangs suggest that a full control requires fewer spectral components for a faithful representation. 
This implies a more convenient situation for Nelder-Mead with a low-dimensional Fourier \crab.
Indeed, the smooth and monotonic controls in Refs.~\cite{doria2011optimal,van2016optimal} also consist roughly of a linear sweep followed by a very steep rise that may be interpreted as the beginning of a bang.  
The results of Refs.~\cite{doria2011optimal,van2016optimal} thus appear consistent with the single bang strategy which is associated with fidelities on the order of mid tens of percent. Given the spread in Fig.~\ref{fig:comparison}(a) such fidelities could reasonably well correspond to the optimized observable values $\rho$ and $\eta$ reported in Refs.~\cite{doria2011optimal,van2016optimal}. 
\\

In conclusion, the presented numerical studies indicate that the exact gradient methodology of Ref.~\cite{jensen2021approximate} is a promising tool in the open-loop setting also for many-body systems.

\section{Unified Optimal Control}
\label{sec:unified}
The results presented in Sec.~\ref{sec:statetransfer} focused solely on ideal open-loop conditions. 
We now discuss the strengths and weaknesses of open- and closed-loop optimization, and how the former can be best utilized in prospective unified frameworks. \\

\textit{Open-loop. ---}
Model-based open-loop methodologies can draw on very efficient derivative-based nonlinear optimization machinery \cite{nocedal2006numerical}
and allows a large degree of numerical parallelization. 
The produced controls are optimal with respect to the chosen mathematical optimization objective $J\st{theory}(\vec u)$ which typically contains contributions from both the quantum process and the experimental constraints.
An optimal control $\vec u^*$ numerically attained through this procedure can therefore only be expected to be as good as the underlying model of the experimental reality, $J\st{experiment}(\vec u^*) \gtrsim J\st{theory}(\vec u^*)$, 
and the acceptance of this potential performance degradation depends on the given context. 
These modeling errors could originate from many sources, for example imperfect equipment fabrication, drifting or fluctuating noisy signals, or nondeterministic run-to-run system preparation. 
If these errors can be characterized as statistical uncertainties, one can endow the optimal controls with robustness towards these by e.g. \textit{ensemble optimization}. 
This places an increased emphasis on both gradient computation speed and exactness.
The exact gradient optimization applied here addresses both these points and is therefore particularly well suited for such extensions, see Ref.~\cite{jensen2021approximate} for a more detailed discussion. 

Nevertheless, if the error sources are unknown and the model cannot be sufficiently refined to an acceptable degree then a $\vec u^*$ found by open-loop control is inadequate.  \\

\textit{Closed-loop. ---}
The primary strength of closed-loop control lies in integrating the experiment itself in the optimization objective.
This model-free approach allows $J\st{experiment}(\vec u)$ to be optimized directly and robustness towards fluctuations is achieved more naturally, given that they are not too large. 
Within this paradigm, gradients can in principle be calculated by finite differences but this is typically too impractical \cite{leng2019robust}.
(This is true regardless of how experimentally accessible the optimized observable is, e.g. fidelity, density of defects, or the rescaled average variance in Sec.~\ref{sec:statetransfer}.)
By instead employing derivative-free direct search approaches, requiring only evaluation of $J\st{experiment}$ itself, 
the numerical programming effort is significantly reduced since it can be outsourced to off-the-shelf ``blackbox'' optimization implementations.   
These approaches are enabled in large by the widely adopted \textsc{crab} technique \cite{muller2021one,caneva2011chopped} discussed in Sec.~\ref{sec:statetransfer}. By choosing an appropriate set of functions, e.g. Fourier components where the expansion coefficients play the role of optimization parameters, the search space is both significantly reduced and focused on certain realistic control shapes. 

However, as also discussed in Sec.~\ref{sec:statetransfer}, derivative-free methods are practically limited in the number of optimization parameters, and this potentially obfuscates the ultimately best implementable solutions.   
Additionally, parallelization is greatly prohibited because the optimization relies on evaluating the cost on (possibly singular) experimental hardware. \\

\textit{Unified-loop. ---}
Open- and closed-loop control thus have distinct and individual advantages and disadvantages. 
The fundamental goal of either approach is the same and combining these in a complementary unified-loop approach appears sensible \cite{egger2014adaptive}. 
Such meta-iterative techniques can be strengthened further by also including model calibration \cite{wittler2021integrated}. 

We argue that an open-loop component is valuable even in the simplest constellation. 
Derivative-based algorithms and their parallelizable nature allow large scale studies and, given that they are sufficiently efficient and accurate, can lead to the discovery of optimal control {strategies} \cite{jensen2020crowdsourcing}.
Identification of such general solution structures and associated physical processes reveals a more complete characterization of a given problem
than any singular solution does, and these insights can be used for choosing appropriate seeds (starting points for the optimization) and parametrizations for Eq.~\eqref{eq:crab}.
Thus, if the modeling is sufficient for correctly representing the optimal \textit{strategy}, it is of much less importance if any \textit{individual} control  degrade in practice because the gap between theory and experiment can be minimized in subsequent closed-loop optimizations and calibrations.

As example, the control strategies found in Sec.~\ref{sec:statetransfer} structure are not \textit{a priori} obvious.
Like in the open-loop case, an isolated derivative-free closed-loop effort using a ``standard'' \cite{muller2021one} low-dimensional chopped Fourier parametrization could not simultaneously capture both the linear and bang-bang-like behavior. 
However, based on the strategy insights achieved by the presented gradient open-loop efforts, one could now, e.g., choose an appropriate reference ramp $u\st{ref}(t)$ and $\{f_l(t;\vec \theta)\}_{l=1}^L$ to be smoothed bangs where the optimization parameters $\vec \theta$ are the widths, positions, and heights of these. 
It was also found that the optimal number of bangs depended on $T$ and this gives an idea of what $L$ should be. 
In addition to subsequent closed-loop optimization in this refined parametrization, the gradient-based open-loop version in Eq.~\eqref{eq:group} may also prove useful in this paradigm.

This broadens the possibility of alternating between open- and closed-loops as e.g. proposed in Refs.~\cite{egger2014adaptive,wittler2021integrated}.

\section{Conclusion}
\label{sec:discussion}
We applied a recently developed quantum optimal control methodology \cite{jensen2021approximate} to a representative example from the class of very
high-dimensional, many-body state transfer problems described by matrix product states where exact diagonalization is prohibited.
Our methodology provides exact gradients by explicitly including in their analytical derivation the effect of 1) Trotterizing the dynamics and 2) choosing a basis where the control Hamiltonian is diagonal. 
This circumvents detrimental scalings with the Hilbert space dimensionality which prohibit the use of other currently known exact gradients in high-dimensional, e.g. many-body, systems. 
The gradient computation time scales instead only with the time it takes to solve the quantum dynamics, i.e. the fundamental operation of any quantum control algorithm. \\

The exact gradient optimizations lead to very high-fidelity results for this class of problems (0.99-0.9999). 
A new type of solution for the $\SF\rightarrow\Mott$ transition was also uncovered, a predominantly linear sweep across the critical point of the phase transition followed by a variable number of bangs (depending on the process duration).   
These solutions remain hidden for the earlier gradient-free approaches \cite{doria2011optimal,van2016optimal} since the employed \crab~ parametrization cannot practically include a sufficient number of Fourier components, i.e. have a high enough bandwidth, to resolve these overall shapes.
A (small) \crab~ parametrization is a practical computational requirement for gradient-free methods, but for gradient-based methods  --- once unencumbered from the aforementioned detrimental scalings --- it is a matter of choice which is free of principal basis size limitations.

Direct quantiative comparisons to \cite{doria2011optimal,van2016optimal} were somewhat hampered by the differences in the specific problem formulations, e.g. choice of figure of merit, but the afforded inferences are very encouraging. 
When the gradient is readily available, gradient-based approaches are generally considered favorable over gradient-free in the idealized open-loop setting
when all else is equal (figure of merit, control parametrization, bandwidth limitation, etc.), and we expect that this would also apply here given a more extensive investigation in a collated numerical environment. \\

The documented efficiency over wide spans of Hilbert space dimensionality (in Ref.~\cite{jensen2021approximate} and here) suggests that the methodology could be useful for future quantum optimization tasks. \\

In discussing the role of open-loop methodologies as a whole, we argued that they are particularly useful both for initially identifying appropriate control subspaces
for a \crab~ parametrization and, in conjunction with subsequent closed-loop methods, may remain a relevant optimization paradigm in the future.

\begin{acknowledgments}
We thank T. Calarco, S. Montangero, K. Mølmer, M. Dalgaard, and C.A. Weidner for feedback.
This work was funded by the ERC, H2020 grant 639560 (MECTRL), and the John Templeton and Carlsberg Foundations.
The numerical results presented in this work were obtained at the Centre for Scientific Computing, Aarhus phys.au.dk/forskning/cscaa.
\end{acknowledgments}

\appendix

\section{Bose-Hubbard model in optical lattice}
\label{app:modelling}
The Bose-Hubbard model in Eq.~\eqref{eq:H} can be realized in a variety of physical platforms \cite{sachdev2011quantumphasetransitions}
and here we review the standard treatment for a cubic optical lattice. 
We first specify how the energetic parameters in Eq.~\eqref{eq:H} relate to the optical trapping depths and the conversion between simulation and laboratory time scales. 
These are then numerically calculated for a set of experimental parameters.


%

We assume a simple cubic periodic lattice approximated near the trap center by 
\begin{align}
V(\vec r ) \approx \sum_{q = x,y,z} v_q \sin^2 k_lq = V_x+ V_y+ V_z,
\end{align}
with $V_q \equiv v_q \sin^2 kq$ being the potential in $q$-direction with depth $v_q$.
Here, $k_l = 2\pi / \lambda_l = \pi/a\st{lat}$ is the laser wavenumber, $\lambda_l$ is the laser wavelength, and $a\st{lat}$ is the lattice site separation. 
Thus, the potential is separable in all directions and we may decompose arbitrary wave functions as $\psi(\vec r) = \psi_x\psi_y \psi_z$. In particular, we may focus on a single direction for the single-particle stationary states. Choosing the $x$-direction, we write $\op H^{\mathrm{1p}}_x \phi_k^n(x) = E_{k}^n \phi_k^n(x)$ where 
$\H^{\mathrm{1p}}_x$ is the single-particle operator in the $x$-direction, $n$ is the band index, $k_l \leq k \leq k_l$ defines the first Brillouin zone of quasi-momentum with intra-equidistant spacing $\Delta k = 2\pi/L = 2\pi /(N_s  a\st{lat})$ where $L$ is the length of the chain. The Bloch wave expansion reads $\phi_k^n(x) = e^{i k x} u_k^n(x)$ where $u_k^n$ inherits the $V$ periodicity. The Fourier series for both quantities contain only a few terms and substitution into the eigenproblem yields a particularly small, simple system of equations for the Fourier expansion coefficients of $u_k^n$ \cite{weidemuller2011interactions}. After numerically obtaining $\phi_k^n(x)$ for a given value of $v_x$, the $n$'th band Wannier state centered on site $i$ is defined by 
\begin{align}
w_{n,x}(x-x_i) = \frac{1}{\sqrt{\mathcal N}} \sum_{k}^{\text{1st Brillouin}} e^{-ikx_i} \phi^n_k(x),
\end{align}
where $x_i = i a\st{lat}$ and $\mathcal N$ is a normalization constant. To progress, 
we make the standard assumptions that the lattice has been loaded in the  ``tube'' of sites defined by $\vec r_i = (x_i,0,0)$ and that
$v_y$ and $v_z$ are sufficiently deep to suppress all tunneling events in their respective directions, and that tunneling along $x$ is nearest neighbor only. Additionally assuming that only the $n=0$ band is occupied in each direction ($v_i \gtrsim 2 E_R$) and dropping the index, the bosonic field operator can be expanded as $\op \Psi(\vec r) \approx \sum_{i=1}^{N_s} \op a_{x_i, 0, 0} \cdot w_x(x-x_i) w_y(y) w_z(z)$. 
Inserting this expansion in the many-body Hamiltonian for a dilute bosonic system \cite{dalfovo1999theory}, one obtains Eq.~\eqref{eq:H} by letting $\op a_i \equiv \op a_{x_i,0,0}$ and defining the constitutive relations 
\begin{align}
\!\!\!J_x (&v_x) = -\int_{-\infty}^{\infty} w_x(x-x_i) \op H^{\mathrm{1p}}_x  w_x(x-x_{i+1}) \d x, \label{eq:Jx} \\
\!\!\!U(&v_x,v_y,v_z) = \g\int|w_x(x-x_i) w_y(y) w_z(z) |^4 \d \vec r \label{eq:U},
\end{align} 
where $\g = 4\pi \hbar^2 a_s / m$ is the two-body collision coupling strength, $a_s=101 a_0$ the $s$-wave scattering length of Rubidium 87, and $a_0$ is the Bohr radius.
Thus, for a cubic optical lattice loaded with ultracold atoms, the energies are related to the trapping depths $v_x,v_y,v_z$ through the Wannier states.
Note the integrals over $y$ and $z$ in $J_x$ are equal to one due to their normalization, and $U$ factors into three independent one-dimensional integrals. 
Both $J_x$ and $U$ nontrivially depend on $v_x$, and although we assume only 1D dynamics along the $x$-direction, the frozen out transverse $y$- and $z$-directions still implicitly enter in $U$ via the associated Wannier functions $w_y$ and $w_z$.
With the constitutive equations in Eqs.~\eqref{eq:Jx}--\eqref{eq:U} we can map any $u_n = U(t_n)/J_x(t_n)$ ramp into the corresponding trapping depth $v_x(t_n)$ as shown below. \\

The natural simulation time scales in Eq.~\eqref{eq:bhcontroldrift}, $\H\st{SI}/J_x = \H$, depends on the control. 
This can be seen by considering the nondimensionalized propagator,
\begin{align}
\U&= \exp\left(-i \frac{\H\st{SI} \dt\st{SI}}{\hbar}\right) = \exp\left(-i \frac{\H\st{SI}}{J_x} \left\{\frac{J_x \mu\st{time}}{\hbar} \right\}\dt\st{sim}\right) \notag\\
& = \exp\left(-i \H \dt\st{sim}\right) \Rightarrow \mu\st{time} = \hbar/J_x,
\end{align}
where time-dependences have been omitted for clarity.
These are working equations corresponding to $\hbar = 1$ and where $\dt\st{SI} =  \mu\st{time} \cdot \dt\st{sim}$. Time steps expressed in SI units, $\dt\st{SI}$, are related to (constant) dimensionless simulation numbers $\dt\st{sim}$, through the time scale $\mu\st{time} = \hbar/ J_x(u_n)$ which depends on the control value.
In particular, the total duration of the transfer process given in SI time is 
\begin{align}
T\st{SI}(\vec u) = \hbar \dt\st{sim}  \sum_{n=1}^{\nt-1}   J_x^{-1}(u_n), \label{eq:T_SI}
\end{align}
i.e. the relevant time scales are given by the specific realization of the physical platform and depends on the control vector. 
Elsewhere in the paper, subscripts are dropped and we write $\dt = \dt\st{sim}$ and all quantities of time are implicitly given in nondimensional simulation values unless followed by a unit. 

\subsection{Experimental parameters}
To enable a degree of quantitative comparison in the main text, we consider the experimental lattice parameters given in Ref.~\cite{van2016optimal} as follows. A summary of our problem specification compared to Refs.~\cite{doria2011optimal,van2016optimal} are shown in Table~\ref{tab:comparison}.

\begin{figure}
	\includegraphics[width=\linewidth]{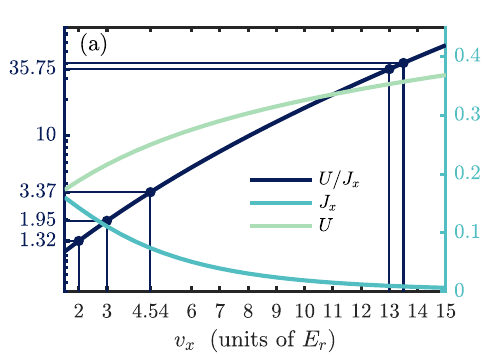}
	\includegraphics[width=\linewidth]{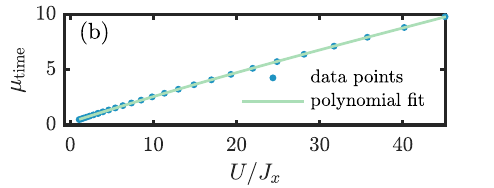}
	\caption{
		(a): Constitutive relations $U(v_x)$, $J_x(v_x)$ for the concrete optical lattice described in the text (right axis in units of $E_R$).  
		The dots denoted on the $U/J_x$ curve (left axis) indicate values of note. Left to right, they are: the minimally allowed $v_x=2\,E_R$, the $v_x$ defining $\SF$, the critical point of the $\SF \rightarrow \Mott$ phase transition, the $v_x$ defining $\Mott$, and the maximally allowed $v_x = 13.5\,E_R$ (corresponding to $U/J=40.18$).
		(b): The time scaling is $\mu\st{time} = \hbar / J_x(v_x)$ in units of ms and the transfer duration in SI units Eq.~\eqref{eq:T_SI} thus depends on the control vector $\vec u = \left( \dots, (U/J_x)_n, \dots \right)$.
	}
	\label{fig:UJ}
\end{figure}

The lattice recoil energy is $E_R = \hbar^2 \pi^2/(2 a\st{lat}^2 m) \approx 2.03 \,\si{kHz} \times h$ where $m=87\, \si{amu}$ is the mass of Rubidium 87 and $a\st{lat}$ is the lattice site separation.  Additionally, $h$ is the Planck constant, and $\hbar$ the reduced Planck constant. We assume a lattice of wavelength $\lambda = 1064\,\si{nm}$ with lattice spacing $a\st{lat} = \lambda/2 = 532\,\si{nm}$.
The transverse trapping depths are fixed at $v_y = v_z = 20 \,E_R$ and the $\SF \rightarrow \Mott$ transition is defined by 
\begin{align}
\ket{\psio} &=  \SF \hspace{0.33cm}\equiv \ket{\mathrm{GS}; v_x = 3 E_R},  \\
\ket{\psitgt} &= \Mott \equiv \ket{\mathrm{GS}; v_x = 13 E_R},
\end{align}
where GS refers to the ground state at the specified longitudinal depth $v_x$. 
The phase transition is driven by varying $v_x$ with the requirement that $v_x  \geq 2\,{E_R}$ at all times to satisfy the modeling assumptions in Eqs.~\eqref{eq:Jx}--\eqref{eq:U}.
With this choice of parameters, the constitutive equations between $v_x$ and the energies $U$ and $J_x$ are calculated numerically and shown in Fig.~\ref{fig:UJ}. The conversion to $U/J_x$ for several relevant depths $v_x$ are shown, 
e.g. at $(v_x)\st{crit} \approx 4.5\, E_R$ that in Ref.~\cite{van2016optimal} corresponds to the critical point for the phase transition.
At this depth we obtain $(U/J_x)\st{crit} \approx 3.4$ which agrees with the number stated in Ref.~\cite{van2016optimal}, and we consider this a verification for our numerical calculation of $J_x(v_x)$ and $U(v_x,v_y,v_z)$.

\begin{widetext}
	\begin{table*}
		\centering
		\addtolength{\tabcolsep}{+2pt}    
		\def\arraystretch{1.2}
		\begin{tabular}{l | l l l}
			\hspace{-0.1cm}& \hspace{0.055cm} This work  &\hspace{0.4cm} Ref.~\cite{doria2011optimal} & \hspace{0.45cm} Ref.~\cite{van2016optimal}\\
			\hline\hline \vspace*{-0.15cm}\\
			\textbf{Physical system} \\
			Sites and particles & $N_s = N_p = 20$ & $N_s=N_p = 20 \text{ (and others)}$ &  $N_s = 32, N_p=16$ \\ 
			Harmonic trapping & No & No/Yes& Yes   \\
			Lattice parameters & 
			\begin{tabular}{@{}l@{}} $\lambda = 1064\,\si{nm}$, \\ $v_y=v_z = 20\,E_R$ \end{tabular} &
			\begin{tabular}{@{}l@{}} $\lambda = 826\,\si{nm}$, \\ $v_y=v_z = 30\,E_R$ \end{tabular} & 
			\begin{tabular}{@{}l@{}} $\lambda = 1064\,\si{nm}$, \\ $v_y=v_z = 20\,E_R$ \end{tabular} \\
			\\
			\textbf{Optimization details} \\
			Objective & $J_F + J_\alpha + J_\gamma$ Eq.~\eqref{eq:totalJ} & 
			\begin{tabular}{@{}l@{}}Energy minimization  halting \\  below $\rho=10^{-3}$ Eq.~\eqref{eq:rho} \end{tabular}
			& $\eta$ Eq.~\eqref{eq:eta} \\
			Update rule & Exact gradients (\bfgs) & Gradient-free (Nelder-Mead) & Gradient-free (Nelder-Mead)\\
			Control & $U/J_x$ & $v_x$ & $v_x$ \\
			Parametrization & 
			None
			& Fourier \crab~  Eq.~\eqref{eq:crab}  & Fourier \crab~ Eq.~\eqref{eq:crab} \\
			Bandwidth limited by & $J_\gamma$ cost & Parametrization &  Parametrization \\
			Time scale & $\hbar/J_x$ & $\hbar/E_R$ & $\hbar/E_R$ \\
			Initial and target depths & $3E_R \rightarrow 13 E_R$ & $2E_R \rightarrow 22 E_R$ & $3E_R \rightarrow 14 E_R$ \\
			\mps~ parameters & $D=200, s\st{max} = 10^{-12}, d=5$ & $D\leq 100$ & $D\leq 24, s\st{max} = 10^{-5}$ \\
			$\dt$ for simulation & $0.025$ & $0.01-0.001$ & $0.01$\\
		\end{tabular} 
		\addtolength{\tabcolsep}{-5pt}    
		\caption
		{
			Summary of problem specifications in this work and Refs.~\cite{doria2011optimal,van2016optimal}. 
			The optimization in this work does not rely on explicit lattice parameters as discussed in Appendix~\ref{app:modelling}. 
			The Refs.~\cite{doria2011optimal,van2016optimal} do not report the concrete details of the control parametrization and certain other parameters for the optimization. 
		}
		\label{tab:comparison}
	\end{table*}
\end{widetext}

\section{Matrix Product States}
\label{app:mps}
We briefly present the many-body ansatz of matrix product states, see Ref.~\cite{schollwock2011density} for an excellent and more detailed introduction.
We then discuss a t-\dmrg algorithm tailored to the necessary problem representation to significantly accelerate computations.

The explosive growth of Hilbert space with the number of constituents is well-known.
This exponential scaling, however, is in a sense a ``convenient illusion'' since the majority of physically relevant states occupy only a small corner of the full Hilbert space \cite{poulin2011quantum}.
These are usually characterized by low entanglement, as measured e.g. by entanglement entropy, and includes ground states and reachable states from these in finite time. 
Tensor networks and their bespoke algorithms are capable of targeting this much reduced subspace with subexponential resources \cite{vidal2003efficient,lloyd2014information,montangero2018introduction}.
The success of such approaches is owed to the fact that the size of the corner is governed by favorable so-called area scaling laws \cite{perez2007matrix}
for the entanglement entropy. 
Matrix product states, also known as tensor trains, are the appropriate types of tensor networks for 1D systems and their area scaling law is constant with the number of constituents.

The general form of a matrix product state for a finite unclosed chain of $N_s$ constituents/sites is
\begin{align}
\ket{\psi} = \sum_{j_1,j_2,\dots,j_{N_s}} \mat A^{j_1} \mat A^{j_2}\cdots \mat A^{j_{N_s}} \ket{j_1,\dots,j_{N_s}}, \label{eq:mps}
\end{align}
where $j_i \in \{1,2,\dots,d\}$ is the \textit{physical index} (degree of freedom) for the $i$'th constituent, $d$ is the size of the local Fock space, and $\mat A^{j_i} \in \mathbb C^{a_{i-1} \times a_{i}}$ where $a_i$ is the \textit{bond index} with the only requirement that the product of all the matrices yields a scalar. The ansatz Eq.~\eqref{eq:mps} is simply a decomposition of 
the expansion coefficient $c_{j_1,j_2,\dots,j_{N_s}}$ tensor of rank $N_s$ into $N_s$ rank 3 tensors $\{A^{j_i}_{a_{i-1},a_i}\}_{i=1}^{N_s}$, which is always possible by repeated singular value decomposition (\textsc{svd}) or similar and is in principle an \textit{exact} representation. Properties of e.g. the \textsc{svd} procedure, however, allows significant truncation of the matrix $\mat A^{j_i}$ dimensions associated with the bond indices for low-entanglement states: singular values of the \textsc{svd} correspond to the expansion coefficients in the Schmidt decomposition across a given bi-partitioning of the system, many of which are close to (or exactly) zero for states in the ``small corner''. Thus, we can choose to keep only singular values larger than a given threshold $s\st{max}$ and/or impose a maximum number values $D$ to keep, depending on the desired accuracy.
Even though the matrix product state is in practice not constructed directly from the coefficients since it requires exponential amounts of storage, virtually all basic matrix product state algorithms such as diagonalization (\dmrg) and time evolution (t-\dmrg) similarly employ \textsc{svd} (or \textsc{qr}) decompositions. 
This enables a natural way of keeping resource consumption in check, typically by specifying a given $s\st{max}$ and/or $D$ in advance.

In the present case of the Bose-Hubbard model Eq.~\eqref{eq:bhcontroldrift} with unit filling,  $j_i$ is the site occupation number and $d=N_p = N_s$ where $N_p$ is the number of particles. The Hilbert space dimension scales exponentially 
\begin{align}
D\st{\mathcal{H}} = \frac{(N_s + N_p - 1)!}{N_p!(N_s - 1)!},
\end{align}
which limits the computational feasibility of exact diagonalization approaches to roughly $N_p = 10-13$ with increasing layers of analytical and numerical sophistication needed for relatively small gains \cite{zhang2010exact}. 
At such a low number of sites, the ``bulk'' of the system is constituted by only a relatively small fraction of sites.
Matrix product states, on the other hand, are associated with polynomial scaling \cite{vidal2003efficient,poulin2011quantum,schollwock2011density} and can comfortably extend this range into the low-to-mid tens of particles in a time-dependent setting \cite{doria2011optimal,van2016optimal,kohn2020superfluid}
or low hundreds in a static setting \cite{ejima2012characterization}.

\subsection{t-DMRG for Bose-Hubbard Model}
Time evolution is the fundamental operation for quantum optimal control. 
For this reason, we present here a t-\dmrg variant similar to that in Ref.~\cite{daley2004time} tailored to the structure of Eq.~\eqref{eq:bhcontroldrift} to speed up our computations. 
\begin{figure}[t]
	\includegraphics[width=\linewidth]{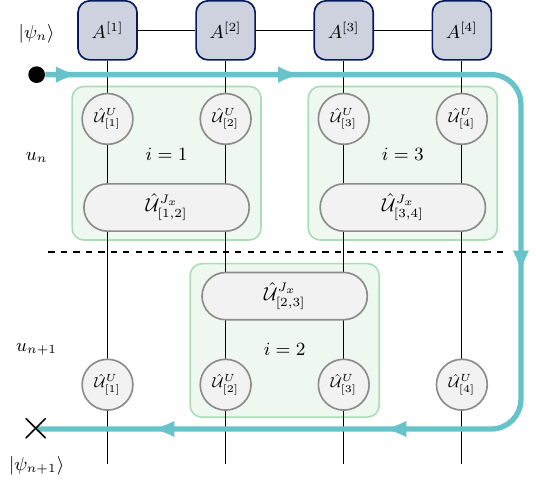}
	\caption{Example tensor network diagram \cite{schollwock2011density} 	
		showing the full calculation $\ket{\psi_{n+1}} = \U_n\ut{ST}\ket{\psi_n} = \U^c_{n+1}\U^d\U^c_n\ket{\psi_n}$ \eqref{eq:fullprop} 
		due to the Hamiltonian in Eq.~\eqref{eq:bhcontroldrift}  for $N_s=4$.}

	\label{fig:ModifiedTE}
\end{figure}

We start by considering the Suzuki-Trotter expansion \cite{jensen2021approximate} in Eqs.~\eqref{eq:ST}.
Each term in the diagonal control Hamiltonian commutes and we may write exactly
\begin{align}
\U^{c/2}_n &= \exp\left({-i  \left(\frac{u_n}{2}\sum_{i=1}^{N_s} \h^U_{[i]}\right)\frac{\dt}{2}}\right) 
= \prod_{i}^{N_s} \U_{n,[i]}^U,
\end{align}
where $\U_{n,[i]}^U = \exp({-iu_n \h^U_{[i]} \dt/4})$.
For the drift Hamiltonian we can apply the same technique as in standard t-\dmrg \cite{schollwock2011density,paeckel2019time} for nearest-neighbor Hamiltonians to obtain a first-order Suzuki-Trotter expansion with associated error $\mathcal{O}(\dt^2)$,
\begin{align}
\U^d &= e^{-i(\H^d\st{even} + \H^d\st{odd} ) \dt} \approx e^{-i\H^d\st{even}  \dt} e^{-i \H^d\st{odd}  \dt} \notag\\
&= \left(\prod_{i\text{ even}}^{N_s-1} \U_{[i,i+1]}^{J_x}  \right)\left(\prod_{i\text{ odd}}^{N_s-1} \U_{[i,i+1]}^{J_x}  \right), \label{eq:bhdriftprop}
\end{align}
where $\U_{[i,i+1]}^{J_x} = \exp(-i\h^{J_x}_{[i,i+1]} \dt)$. The enabling step in this expansion is to group even and odd terms
\begin{align}
\H^d &= \H^d\st{even} + \H^d\st{odd} = \sum_{i \text{ even}}^{N_s-1} \h^{J_x}_{[i,i+1]} + \sum_{i \text{ odd}}^{N_s-1}\h^{J_x}_{[i,i+1]}.
\end{align}
Although $[\H^d\st{even}, \H^d\st{odd} ] \neq 0$ causes the $\mathcal{O}(\dt^2)$ error, each term has total internal-commutativity, allowing the subsequent exact product form Eq.~\eqref{eq:bhdriftprop}.
Combining the above expressions and moving each individual even (odd) $\h^{J_x}_{[i,i+1]}$ to the left (right) until they meet a noncommutative operator, we obtain for even $N_s$
\begin{align}
\U\ut{ST}_n 
\approx &
\prod_{i}^{N_s} \U_{n+1,[i]}^U
\prod_{i\text{ even}}^{N_s-1} \U_{[i,i+1]}^{J_x}
\prod_{i\text{ odd}}^{N_s-1} \U_{[i,i+1]}^{J_x}
\prod_{i}^{N_s} \U_{n,[i]}^U
\notag \\
= \;
&\U_{n+1,[1]}^U\left( 
\prod_{N_s-1}^{i\text{ even}}
\U_{n+1,[i]}^U \U_{n+1,[i+1]}^U
\U_{[i,i+1]}^{J_x} 
\right)  \notag
\\
\times \;&\U_{n+1,[N_s]}^U
\left(
\prod_{i\text{ odd}}^{N_s-1} \U_{[i,i+1]}^{J_x}   \U_{n,[i+1]}^U \U_{n,[i]}^U
\right) \notag
\\
\equiv\;
&\U_{n+1,[1]}^U\left( 
\prod_{N_s-1}^{i\text{ even}}
\U_{n+1,[i,i+1]}^{UUJ_x}
\right) \label{eq:backsweep} \\
\times \;&\U_{n+1,[N_s]}^U 
\left(
\prod_{i\text{ odd}}^{N_s-1} \U^{J_x UU}_{n,[i,i+1]}
\right) \label{eq:frontsweep}  \\
\equiv\; &\U_{n+1,[1]}^U \,\U\st{backsweep}\,\U_{n+1,[N_s]}^U\U\st{forwardsweep}.  \label{eq:fullprop}
\end{align}
If $N_s$ is odd, replace $\U_{n+1,[N_s]}^U \rightarrow \U_{n,[N_s]}^U$ in the final expressions.
In the language of matrix product states, application of one-site $\U_{n,[i]}^U$ gates and two-site gates $\U_{[i,i+1]}^{J_x}$ can be done very efficiently when exploiting left- and right-normalization of the site tensors. 
The one-site gates are particularly cheap to compute because $\h^U_{[i]}$ is diagonal. 
The two-site gates, which would otherwise entail the most expensive operation, are time-independent and can be precomputed, stored on the disk, and be loaded into memory on runtime.
Additionally, the grouping of product triples, e.g. $\U_{n,[i,i+1]}^{J_xUU}$, acting only on nearest-neighbor pairs of indices $[i,i+1]$ provides a way of reducing overhead in the tensor network contraction $\U\ut{ST}_n \ket{\psi}$ by advancing the central site (gauge) of the matrix product state: 
apply the product of triples and contract the site tensors in a ``forward sweep'' over odd  $i$ \eqref{eq:frontsweep} and then
in a ``backward sweep'' over even $i$ \eqref{eq:backsweep} as illustrated in Fig.~\ref{fig:ModifiedTE}. A more technical detailing is as follows.

The site tensors (purple nodes) are connected by auxiliary/bond indices (black solid horizontal lines). 
The one- and two-site gates (gray nodes, time index suppressed) come in triples (green boxes) and are applied to physical indices of the site tensors (black solid vertical lines), in the order indicated by the thick teal arrowed line according to the sweeping order. The beginning (end) is marked by a black dot (cross), and the dashed line separates the forward sweep with $u_n$ (above the line) from the backward sweep with  $u_{n+1}$ (below the line). 
Application of product triple $i$ in the forward sweep entails the following: 
\begin{enumerate}
	\item[(1)] Contract the two site tensors with physical indices $i$ and $i+1$ over their common bond index
	into a temporary two-site tensor.
	\item[(2)] Apply the two one-site gates followed by the two-site gate.
	\item[(3)] Split the temporary two-site tensor by \textsc{svd} back into two individual site tensors with the central site (gauge) moved to $i+1$.
	\item[(4)] Shift the gauge by \textsc{svd} an additional site to the right such that the central site is at $i+2$.
\end{enumerate}
Each arrow tip demarcates the gauge position during the sweeps, where sites to the right (left) are right(left)-normalized, with the exception that the next site intersecting the orange line is the central site.
After applying the first left-over one-site gate, the backward sweep is similarly performed with the following modifications: (1) the site indices are $i-1$ and $i$,  (2) the order of gate application reversed, (3) the central site is placed on $i-1$, and (4) the central site is gauged to $i-2$. Finally, the second left-over one-site gate is applied and this completes the time step.
The same procedure with new control values $u_{n+1}$ and ${u_{n+2}}$ can subsequently be applied to obtain $\ket{\psi_{n+2}} = \U_{n+1}\ut{ST}\ket{\psi_{n+1}}$.   	  
Implementing the backward propagation $\ket{\psi_{n}} = \U_n\ut{ST \dagger}\ket{\psi_{n+1}}$ is similar, but with reversed arrow tips and order of application.

\section{Optimization Details}
\label{app:opt}
\begin{figure}[t]
	\includegraphics[]{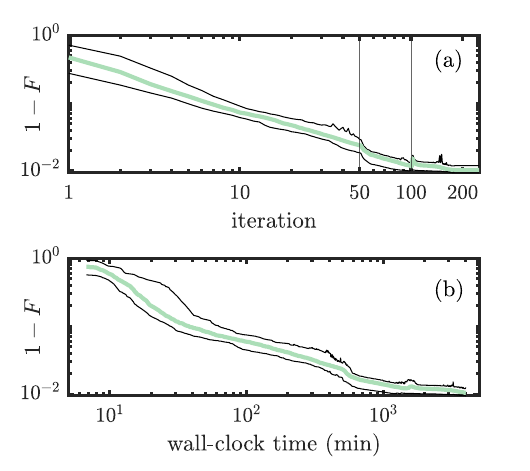}
	\caption{Optimization trajectories  $(5\%,50\%,95\%)$-quantiles for $1-F$ as a function of (a): iteration and (b): optimization wall-clock time for the  $100$ seeds at $T=11$ in Fig.~\ref{fig:FT99}(a). The vertical lines mark changes in the $\dt$ homotopy parameter.
	} 
	\label{fig:traces}
	\vspace{0.5cm}
	\includegraphics[]{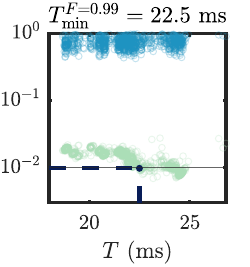}
	\includegraphics[]{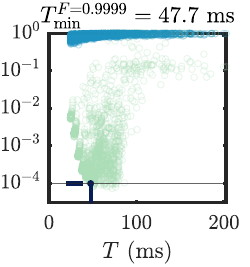}\\
	\caption{The $1-F(T)$ results from Fig.~\ref{fig:FT99} plotted against SI instead of simulation units.
		\label{fig:FT_SI}
	}
\end{figure}

Our matrix product state computations are performed using the \textsc{itensor} library \cite{ITensor}. We use an auxiliary dimension of $D=200$, a singular value threshold of $s\st{max} = 10^{-12}$, and a reduced local Fock space $d=5$ 
(higher local occupation numbers do not contribute significantly to the dynamics due to the exponential on-site energy penalty).
For reference, the benchmark Refs.~\cite{doria2011optimal,van2016optimal} used $D\leq 100$ and $D\leq 24$, $s\st{max} = 10^{-5}$, respectively, corresponding to less computationally expensive, more approximate, low-entanglement representations of the model.
We use the \dmrg algorithm implemented in \textsc{itensor} to obtain the initial- and target states.

For the given system size of $N_s = N_p = 20$, the durations required to approach the minimal duration for fidelity $F=0.99$, $\tqsl[0.99]$, with sufficiently low Trotterization error necessitates about $\nt = 350-450$ time steps for time steps of size $\dt = 0.025$. 
To accelerate the optimizations we take $\dt$ to be a homotopy/continuation parameter \cite{jensen2021approximate}: we sequentially optimize on increasingly fine grained time grids, specifically $\dt = 0.1 \rightarrow 0.05 \rightarrow 0.025$. By halving the values, the new grid points coincide with the old but with doubled resolution 
as each newly inserted point is set to the value of old point immediately prior corresponding to $\U_n(\dt) \approx \U_n(\dt/2) \U_n(\dt/2)$, where $\U_n$ is the time evolution operator at time index $n$. 
The benefit is that the coarser optimizations can yield relatively rapid fidelity improvements since fine grained resolution is typically not needed for the overall shape of the solution.
Care should be taken not to spend too much time on these, since they are not fully coincidental with the final optimization landscape \cite{jensen2021approximate}. 
Note that this technique is enabled by the exactness of the Trotterized gradient not being dependent on $\dt$ which is not the case for the exact propagator gradient with finite summation cutoffs, see Ref.~\cite{jensen2021approximate}. We stress that the exact derivative in Eq.~\eqref{eq:exactGradientST} is the main workhorse whereas the homotopy and time evolution in Appendix~\ref{app:mps} are secondary but effective acceleration techniques.

The SI time scaling in Eq.~\eqref{eq:T_SI} depends on $J_x^{-1}$ and thus the control value. Figure~\ref{fig:UJ} shows that larger $U/J_x$ values correspond to longer SI times. 
Since we desire the fastest possible optimal controls in real time, we place an upper bound corresponding to $v_x \leq 13.5 \,E_R$ during optimization to limit this artifact of the nondimensionalization. We also add slight preference towards lower control values by introducing a regularization cost term, $J_\alpha$, for the control amplitude [Eq.~(A28) in Ref.~\cite{jensen2021approximate}]. Due to the limited bandwidth of experimental electronics we also add a regularization cost term, $J_\gamma$, for the temporal derivative of the control, shifting preference towards smoother controls [Eq.~(A30) in Ref.~\cite{jensen2021approximate}].
The strength of these terms are controlled by the parameters $\alpha, \gamma \geq 0$, respectively, and typically $\alpha, \gamma \sim 10^{-7}-10^{-10}$.
The total optimization objective is thus
\begin{align}
J(\vec u)=J_F(\vec u) + J_\alpha(\vec u) + J_\gamma(\vec u). \label{eq:totalJ}
\end{align}
The derivatives for these cost terms are calculated in the Appendix of Ref.~\cite{jensen2021approximate} and is included in the optimization.

For the optimization (i.e. search direction and step size line searching in Eq.~\eqref{eq:localupdate}), we employ the nonlinear interior-point algorithm implemented in \textsc{ipopt} \cite{wachter2006implementation} by supplying the exact derivatives. 
Briefly, interior-point methods can handle control constraints by including them explicitly when solving for the searching direction, which in our case is $1.32 \leq u_n \leq 40.18$ for all time indices $n$. Being a second-order derivative method, the search direction includes the Hessian or a gradient-based approximation thereof (\textsc{bfgs}). We found the exact Hessian calculation (time scale of days per iteration) for the problem under consideration to be outside our time budget even when including the homotopy, and therefore opted for the \textsc{bfgs} approach (time scale of hours per iteration). 
The seeds (initial points for the optimization) were optimized in parallel on individual cores in a computer cluster. The results reported in the main text were allotted roughly three to seven days of optimization time.
Our seeding strategy is based on an exponential reference control overlaid with a sum of random Fourier components. 
As a verification for our implementation of e.g. the exact analytical derivatives and time evolution, we compared the analytical derivatives to their finite difference counterpart and found that they agreed to the same precision as in Ref.~\cite{jensen2021approximate}.

Figure~\ref{fig:traces} shows optimization trajectory statistics.
Changes in the $\dt$ homotopy parameter manifest as kinks at 50 and 100 iterations due to $\dt$ changes $0.1 \rightarrow 0.05$, 	and $0.05 \rightarrow 0.025$. A dip in infidelity is seen at the first handover as the increased time resolution of the control allows more complex and fine-tuned dynamics.
The homotopy approach accelerates the computations and roughly doubles the number of iterations achieved within the allocated time budget without sacrificing performance since the infidelity iteration trajectories follow roughly the same exponential-law for all three homotopy parameter regions.

Figure~\ref{fig:FT_SI} shows the results Fig.~\ref{fig:FT99} plotted against their SI duration using Eq.~\eqref{eq:T_SI}.

\bibliography{/Users/au446513/projects/bibmaster/references.bib}

\end{document}